\documentclass[%
reprint,
 citeautoscript,
twocolumn,
 amsmath,amssymb,
 longbibliography,
aip,
floatfix,
]{revtex4-2}

\usepackage{amsmath}
\usepackage{amsthm}
\usepackage{amsfonts}
\usepackage{amssymb}
\usepackage{graphicx}
\usepackage{numprint}
\usepackage{mathtools}
\usepackage[colorlinks=true,linkcolor=blue,citecolor=blue,urlcolor=cyan,bookmarks,breaklinks=true]{hyperref}
\usepackage{enumitem}
\usepackage{graphicx}
\usepackage{dcolumn}
\usepackage{bm}
\usepackage{xcolor}
\usepackage[font=small,labelfont=bf,
   justification=raggedright,
   format=plain]{caption}
\usepackage{subcaption}
\usepackage{subcaption}

\begin{document}
\title{Realizability of Iso-$g_2$ Processes via Effective Pair Interactions}

\begin{abstract}
 An outstanding problem in statistical mechanics is the determination of whether prescribed functional forms of the pair correlation function $g_2(r)$ [or equivalently, structure factor $S(k)$] at some number density $\rho$ can be achieved by $d$-dimensional many-body systems. The Zhang-Torquato conjecture states that any realizable set of pair statistics, whether from a nonequilibrium or equilibrium system, can be achieved by equilibrium systems involving up to two-body interactions.
 In light of this conjecture, we study the realizability problem of the nonequilibrium iso-$g_2$ process, i.e., the determination of density-dependent effective potentials that yield equilibrium states in which $g_2$ remains invariant for a positive range of densities.
 Using a precise inverse methodology that determines effective potentials that match hypothesized functional forms of $g_2(r)$ for all $r$ and $S(k)$ for all $k$, we show that the unit-step function $g_2$, which is the zero-density limit of the hard-sphere potential, is remarkably numerically realizable up to the packing fraction $\phi=0.49$ for $d=1$. For $d=2$ and 3, it is realizable up to the maximum ``terminal'' packing fraction $\phi_c=1/2^d$, at which the systems are hyperuniform, implying that the explicitly known necessary conditions for realizability are sufficient up through $\phi_c$.
 For $\phi$ near but below $\phi_c$, the large-$r$ behaviors of the effective potentials are given exactly by the functional forms $\exp[-\kappa(\phi) r]$ for $d=1$, $r^{-1/2}\exp[-\kappa(\phi) r]$ for $d=2$, and $r^{-1}\exp[-\kappa(\phi) r]$ (Yukawa form) for $d=3$, where $\kappa^{-1}(\phi)$ is a screening length, and for $\phi=\phi_c$, the potentials at large $r$ are given by the pure Coulomb forms in the respective dimensions, as predicted by Torquato and Stillinger [\textit{Phys. Rev. E}, 68, 041113 1-25 (2003)].
 We also find that the effective potential for the pair statistics of the 3D ``ghost'' random sequential addition at the saturation density $\phi_c=1/8$ is much shorter-ranged than that for the 3D unit-step-function $g_2$ at $\phi_c$, and thus does not constrain the realizability of the unit-step-function $g_2$.
 Our inverse methodology yields effective potentials for realizable targets, and as expected, it does not reach convergence for a target that is known to be non-realizable, despite the fact that it satisfies all known explicit necessary conditions.
Our findings demonstrate that exploring the iso-$g_2$ process via our inverse methodology is an effective and robust means to tackle the realizability problem and is expected to facilitate the design of novel nanoparticle systems with density-dependent effective potentials, including exotic hyperuniform states of matter.
    
\end{abstract}
\author{Haina Wang}
\author{Frank H. Stillinger}
\affiliation{\emph{Department of Chemistry, Princeton University}, Princeton, New Jersey, 08544, USA}
\author{Salvatore Torquato}
\email[]{Email: torquato@princeton.edu}
\affiliation{Department of Chemistry, Department of Physics, Princeton Institute of Materials, and Program in Applied and Computational Mathematics, Princeton University, Princeton, New Jersey 08544, USA}
\affiliation{School of Natural Sciences, Institute for Advanced Study, 1 Einstein Drive, Princeton, NJ 08540, USA}

\date{\today}
\maketitle
\newpage
\section{Introduction}
\label{intro}
The relationship between interactions in many-body systems and their corresponding structural properties is fundamental in statistical physics, condensed-matter physics, chemistry, mathematics and materials science \cite{Va06,Re06a,To09a,Co09}. Among the commonly used structural descriptors \cite{To02a}, the pair correlation function $g_2(\textbf{r})$ has unique importance due to its computational simplicity, experimental accessibility through diffraction measurements and the popularity of theoretical many-body models with pairwise additive interactions \cite{To02a,Ha86}. Despite the widespread application of $g_2(\mathbf{r})$, the realizability of pair correlation functional forms by actual many-body configurations remains an outstanding problem in statistical mechanics \cite{Ya61,Cr03,Cos04,Uc06a,To06b,Ku07,Zh20}.
It is known that for a statistically homogeneous (i.e., translationally invariant) many-body system in $d$-dimensional Euclidean space $\mathbb{R}^d$, knowledge of one- and two-body correlations are insufficient to determine the corresponding higher-body correlation functions $g_3,g_4,...$ \cite{To06b}.
Consequently, given a prescribed functional form of a pair correlation function $g_{2}(\textbf{r})$ [or equivalently, a target structure factor $S(\textbf{k})$] and number density $\rho\geq 0$, an uncountable number of necessary and sufficient conditions must be satisfied for $g_{2}(\mathbf{r})$ to be realizable  \cite{Cos04}, which are practically impossible to check. The ensemble-averaged structure factor $S(\mathbf{k})$ is defined as
	\begin{equation}
		S(\mathbf{k})=1+\rho \tilde{h}(\mathbf{k}),
		\label{skdef}
	\end{equation}
where $h(\mathbf{r})=g_2(\mathbf{r})-1$ is the total correlation function, and $\tilde{h}(\mathbf{k})$ is the Fourier transform of $h(\mathbf{r})$. 
While some necessary conditions can be readily checked \cite{Ya61,Cos04,Uc06a}, it remains a practically important task to probe realizability via precise numerical techniques.

A major focus of this study is the realizability of the unit-step pair correlation function for positive densities:
\begin{equation}
    g_{2}(r)=
    \begin{cases}
    0, \quad r \leq D\\
    1, \quad r > D,
    \end{cases}
    \label{g2-us}
\end{equation}
where $D$ is the sphere diameter, hence taken to be unity. Equation (\ref{g2-us}) implies that the many-body system forms a \textit{packing} in which no particles overlap. The \textit{packing fraction}, i.e., the fraction of the total volume covered by the spheres is given by $\phi=\rho v_1(1/2)$, where $v_1(R)$ is the volume of a $d$-dimensional sphere of radius $R$:
\begin{equation}
    v_1(R)=\frac{\pi^{d/2}R^d}{\Gamma(1+d/2)}.
\end{equation}
Importantly, Eq. (\ref{g2-us}) is exactly the zero-density limit of $g_2(r)$ for the equilibrium hard-sphere (HS) fluid \cite{Ha86,To02a}, whose pair potential is given by
\begin{equation}
    v_{\text{HS}}(r)=
    \begin{cases}
    +\infty, \quad r \leq 1\\
    0, \quad r > 1.
    \end{cases} 
    \label{HS}
\end{equation}
Hence, it is not immediately obvious that the step-function $g_2$ given by (\ref{g2-us}) can be achieved for positive densities.

\begin{figure*}[htp]
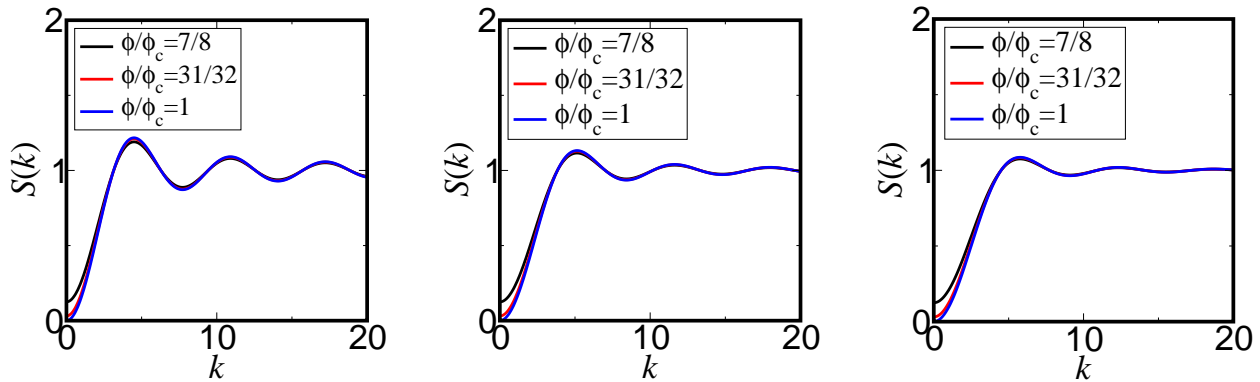

  \centering
  \begin{subfigure}{5cm}
    \includegraphics[width=50mm]{Fig1a.eps}
    \phantomcaption
    \label{1a}
  \end{subfigure}
  \qquad
  \begin{subfigure}{5cm}
    \includegraphics[width=50mm]{Fig1b.eps}
    \phantomcaption
    \label{1b}  
  \end{subfigure}
  \qquad
  \begin{subfigure}{5cm}
    \includegraphics[width=50mm]{Fig1c.eps}
    \phantomcaption
    \label{1c}  
  \end{subfigure}
\captionsetup{subrefformat=parens}
  \caption{Target structure factors for the unit-step-function $g_2$ at various relative packing fractions $\phi/\phi_c$, where $\phi_c=1/2^d$ is the terminal packing fraction, for \subref{1a} 1D, \subref{1b} 2D and \subref{1c} 3D systems.}
  \label{fig:unitstep_S}
\end{figure*}

The $g_2$-invariant process introduced by Torquato and Stillinger \cite{To02c} aims to determine the nonvanishing density range over which a prescribed form of $g_2(r)$ for a many-body system remains invariant over that range. They showed that a $g_2$-invariant process possesses an upper \textit{terminal density} $\rho_c$, which is the highest value such that all explicitly known necessary conditions are satisfied, including $g_2(r)\ge 0$ for all $r$, $S(k) \ge 0$ for all $k$ and the Yamada condition \cite{Ya61,Cos04,Uc06a}, as detailed in Sec. \ref{nec}. The terminal packing fraction in the case of the step function (\ref{g2-us}) is given by \cite{To02c}
\begin{equation}
    \phi_c=\frac{1}{2^d},
\end{equation}
at which the structure factor associated with (\ref{g2-us}) satisfies $\lim_{k\rightarrow 0} S(\mathbf{k})=0$, implying that the system must be hyperuniform \cite{To03a}, if realizable \cite{St05}. 
Disordered hyperuniform systems anomalously suppress large-scale density fluctuations compared to typical disordered systems, such as liquids \cite{To03a}; see Sec. \ref{hu} for detailed definitions. Such exotic amorphous states of matter have been receiving great attention because they connect a variety of seemingly unrelated systems that arise in physics, chemistry, materials science, mathematics, and biology \cite{To18a}. 
Figure \ref{fig:unitstep_S} shows the target $S(k)$ corresponding to the unit-step-function $g_2$ at various densities for $d=1,2,3$ \cite{St05}; see Sec. \ref{ps} for their explicit functional forms.

Previous numerical studies have shown that the unit-step-function $g_2$ is realizable, but only for a finite range of $r$, in one \cite{St01a,Cr03}, two \cite{Cr03} and three \cite{Uc06a} dimensions up to the terminal packing fraction. However, the reverse Monte-Carlo technique \cite{Mc01} used to achieve such realizations does not target small-$k$ correlations in Fourier space and provide no information on interparticle interactions. For $d=1$, Costin and Lebowitz proved that the unit-step-function $g_2$ is realizable by a special type of Markovian point processes (known as renewal processes) if and only if $\phi\leq e^{-1}\sim 0.370$ \cite{Cos04}. 
While the results for this special $g_2$-invariant process do not necessarily mean that the one-dimensional (1D) unit-step-function $g_2$ is non-realizable by \textit{any} process for $e^{-1}<\phi < \phi_c=1/2$, they nevertheless raise the possibility that this 1D target may not be realizable at $\phi_c$.

A powerful way to tackle the realizability problem is via inverse statistical mechanics, i.e., determining the effective interactions that attain an equilibrium system with the prescribed pair statistics \cite{Le85, Ly95, So96, To09a, Jain13, He18, Zh20}.
Recently, Zhang and Torquato conjectured that any realizable  $g_2(\textbf{r})$ or $S(\textbf{k})$ corresponding to a translationally invariant nonequilibrium system can be attained by an equilibrium ensemble involving only (up to) effective pair interactions \cite{Zh20}. 
They introduced a theoretical formalism that enabled them to devise an efficient algorithm to construct systematically canonical-ensemble particle configurations with such targeted $S(\mathbf{k})$ at all wavenumbers, whenever realizable. However, the procedure presented in Ref. \citenum{Zh20} does not provide the explicit functional forms of the underlying one- and two-body potentials. 

Very recently, Torquato and Wang tested the Zhang-Torquato conjecture for challenging target pair statistics that correspond to nonequilibrium systems of current interest \cite{To22, Wa22b}, including a nonhyperuniform 2D random sequential addition process \cite{Fe80,Co88,To06a,Zh13b}, as well as exotic disordered hyperuniform states, such as a 2D perfect glass \cite{Zh16a, Wa22b}, a 3D ``cloaked'' uniformly randomized lattice \cite{Kl20, To22} and a 3D critical-absorbing state \cite{Co08, He15, Wa22b}. 
In all cases, the nonequilibrium target pair statistics can be achieved accurately by equilibrium states with effective one- and two-body potentials, lending crucial support to the Zhang-Torquato conjecture. Further testing this conjecture requires the solution of inverse problems for pair statistics with prescribed functional forms.

In this work, we use a precise inverse methodology that we introduced recently \cite{To22} to numerically investigate the realizability problem via effective interactions. 
Specifically, we study the problem as an \textit{iso-$g_2$ process} \cite{Sa02,St01a,St05}, which is a process corresponding to a many-body system under a density-dependent effective potential $v(r;\rho)$ that yields equilibrium states in which $g_2$ remains invariant for a positive range of densities at constant positive temperature $T$. 
In other words, the effective potential negates the natural density dependence of $g_2(\mathbf{r})$. 
Density-dependent potentials are widely applied to describe coarse-grained models for polymers and macromolecules \cite{st02}. 
The iso-$g_2$ process is a nonequilibrium process whenever the target $g_2$ is known to correspond to an equilibrium states: 
One then studies whether the same equilibrium $g_2$ can be achieved for a range of nonvanishing densities.
Since at positive densities, $g_2(r)$ for the pure equilibrium hard-sphere fluid deviates from the unit-step functional form (\ref{g2-us}) and becomes oscillatory \cite{Ha86}, the corresponding iso-$g_2$ process determines an effective potential $v(r;\phi)$ that suppresses such oscillations (both positive and negative correlations) and maintains the unit-step form (\ref{g2-us}).  
Approximations of the effective potential have been obtained via Percus-Yevick and hypernetted chain (HNC) approximations \cite{St01a}. 
Stillinger and Torquato also derived low-density approximations for the effective potential using diagrammatic expansions \cite{St05}. 
However, these approximate potentials yield $g_2(r)$ that deviate from (\ref{g2-us}), and the deviation increases for larger $\phi$. Thus, these results do not provide a conclusion on the realizability of the unit-step-function $g_2$ for $\phi$ close to $\phi_c$.

Prior to the development of our inverse methodology \cite{To22}, predictor-corrector methods \cite{Le85,Ly95,So96,He18}, such as Iterative Boltzmann inversion (IBI) \cite{So96} and iterative HNC inversion (IHNCI) \cite{Le84, He18}, were regarded to be the most accurate inverse procedures.
Both IBI and IHNCI begin with an initial discretized (binned) approximation of a trial pair potential.
The trial pair potential at each binned distance is iteratively updated to attempt to reduce the difference between the target and trial pair statistics. 
However, IBI and IHNCI cannot treat long-ranged pair interactions required for hyperuniform targets, nor do they consider one-body interactions that stabilize hyperuniform equilibrium states \cite{To22}; see Sec. \ref{hu} for details.
These algorithms also accumulate random errors in the binned potentials due to simulation errors in the trial pair statistics, and thus do not achieve the precision required to probe realizability problems.
Moreover, because all previous methods do not optimize a pair-statistic ``distance'' functional, they are unable to detect poor agreement between the target and trial pair statistics that may arise as the simulation evolves, leading to increasingly inaccurate corresponding trial potentials, as demonstrated in Ref. \citenum{To22}.

Our inverse methodology \cite{To22} improves on previous procedures in several significant ways. 
It utilizes a parameterized family of pointwise basis functions for the potential function at $T>0$, whose initial form is informed by small- and large-distance behaviors dictated by statistical-mechanical theory.
Pointwise potential functions do not suffer from the accumulation of random errors during a simulation, resulting in more accurate interactions \cite{To22}.
Subsequently, a nonlinear optimization technique is utilized to minimize an objective function that incorporates \textit{both} the target pair correlation function $g_2({\bf r})$ and structure factor $S({\bf k})$ so that both the small- and large-distance correlations are very accurately captured. 
For hyperuniform targets, our methodology is able to optimize the required long-ranged pair potential \cite{To08a} as well as the neutralizing background one-body potential \cite{To22}; see Sec. \ref{meth} for details.

To assess the accuracy of inverse methodologies to target pair statistics, we introduced \cite{To22}
the following dimensionless $L_2$-norm error:
\begin{equation}
{\cal E}= \sqrt{D_{g_2}+D_{S}},
\label{L2}
\end{equation}
where $D_{g_2}$ and $D_S$ are $L_2$ functions, given by
\begin{equation}
D_{g_2}=\rho\int_{\mathbb{R}^d} [g_{2,T}(\mathbf{r})-g_{2,F}(\mathbf{r};\mathbf{a})]^2 d\mathbf{r}, 
\label{g2-norm}
\end{equation}
\begin{equation}
D_S=\frac{1}{\rho (2\pi)^d}\int_{\mathbb{R}^d} [S_{T}(\mathbf{k})-S_{F}(\mathbf{k};\mathbf{a})]^2 d\mathbf{k},
\label{S-norm}
\end{equation}
where $g_{2,F}(\mathbf{r};\mathbf{a})$ and  $S_{F}(\mathbf{k};\mathbf{a})$ represent the final
pair statistics at the end of the optimization, which depend on the supervector $\bf a$.
We have previously shown that our inverse methodology generally yields $L_2$-norm errors that are an order of magnitude smaller than those via previous inverse methods, and is able to treat challenging near-critical and hyperuniform targets  \cite{To22}, which previous methods \cite{Le85,Ly95,So96,He18} cannot do. 
Importantly, for equilibrium target pair statistics, it reaches the precision required to recover the unique potential dictated by Henderson's theorem \cite{He74}.
Thus, it is a superior method for our purpose of probing the realizability of pair statistics over all distances in direct (physical) space and all wavenumbers in Fourier space, especially as it concerns hyperuniform targets.

A major goal of this study is to ascertain the density range on which the unit-step-function $g_2$ is realizable for $d=1,2,3$ in the thermodynamic limit and to determine the effective potential when it is realizable. 
We find that for $d=1$, the unit-step-function $g_2$ is numerically realizable up to $\phi=0.49=0.98\phi_c$. 
For $d=2,3$, it is realizable up to $\phi=\phi_c$, at which the systems are hyperuniform.
The effective potentials at the maximum realizable packing fractions of the unit-step-function $g_2$ significantly suppress both positive and negative correlations that would otherwise be present in the pure equilibrium HS system, as manifested by the smaller order metrics $\tau$ (\ref{tau}) for the corresponding former systems.
For $\phi$ near but below $\phi_c$, the large-$r$ behavior of the effective potentials are $\exp[-\kappa(\phi) r]$, $r^{-1/2}\exp[-\kappa(\phi) r]$ and $r^{-1}\exp[-\kappa(\phi) r]$ for $d=1, 2$ and $3$, respectively, where $\kappa(\phi)$ is the inverse screening length, and for the 2D and 3D targets at $\phi=\phi_c$, the potentials at large $r$ are given by the pure Coulomb forms in the respective dimensions, as predicted by Torquato and Stillinger \cite{To03a}. 
This capacity to generate hyperuniform states via interacting many-particle systems is an important advance, since it is expected to facilitate the self-assembly of tunable hyperuniform soft-matter materials and enables one to probe the thermodynamic and dynamic properties of such exotic states \cite{Wa22b}. We also study the realizability of two other targets of particular theoretical importance. The ghost random sequential addition (RSA) process \cite{To06a} is a generalization of the standard RSA process \cite{Re63, wi66, To02a} and is closely related to the unit-step iso-$g_2$ process, as its saturation packing fractions for all $d$ are identical to $\phi_c=1/2^d$. 
At the saturation density,  $g_2$ for the ghost RSA process contains a ``bump'' at the contact radius $r=1$, which may suggests that low-dimensional unit-step-function $g_2$ is unrealizable \cite{To06a}. 
We show that the 3D ghost RSA target at $\phi_c=1/8$ is realizable, and its corresponding effective potential is much shorter ranged than that for the unit-step-function $g_2$. 
We also test the accuracy and power of our inverse methodology for a known non-realizable case that meets all of the known necessary conditions, and yet is not realizable due to geometric constraints \cite{To06b}. Our method is shown to be robust, as it does not reach convergence, as expected, for this known unrealizable target.


We begin by providing basic definitions and background in Sec. \ref{def}. In Sec. \ref{meth}, we describe our inverse methodology to determine effective potentials for targeted pair statistics. Section \ref{unitstep} presents the realizability range and the effective interactions for the unit-step-function $g_2$ for $d=1,2,3$. Sections \ref{ghost} and \ref{impossible} present the results for the 3D ghost RSA target and the 2D non-realizable target, respectively. We provide concluding remarks in Sec. \ref{conclusions}.

\section{Definitions and preliminaries}
\label{def}
\subsection{Pair statistics}
\label{ps}
We consider many-particle systems in $\mathbb{R}^d$ that are completely statistically characterized by the $n$-particle probability density functions $\rho_n(\mathbf{r}_1,...,\mathbf{r}_n)$ for all $n\geq 1$ \cite{Ha86}. In the case of statistically homogeneous systems, $\rho_1(\mathbf{r}_1)=\rho$ and $\rho_2(\mathbf{r}_1,\mathbf{r}_2)=\rho^2 g_2(\mathbf{r})$, $\rho$ is the number
density in the thermodynamic limit, $g_2(\mathbf{r})$ is the pair correlation function, and $\mathbf{r}=\mathbf{r}_2-\mathbf{r}_1$.
If the system is also statistically isotropic, then $g_2(\mathbf{r})$ is the radial function $g_2(r)$, where $r=|\mathbf{r}|$. 

For a single periodic configuration containing $N$ point particles at positions ${\bf r}_1,{\bf r}_2,\ldots,{\bf r}_N$ within a fundamental cell $F$ of a lattice $\Lambda$, the {\it scattering intensity} $\mathcal{I}(\mathbf{k})$  is defined as
	\begin{equation}
		\mathcal{I}(\mathbf{k})=\frac{\left|\sum_{i=1}^{N}e^{-i\mathbf{k}\cdot\mathbf{r}_i}\right|^2}{N}.
		\label{scattering}
	\end{equation}
For an ensemble of periodic configurations of $N$ particles within the fundamental cell $F$, the ensemble average of the scattering intensity in the infinite-volume limit is directly related to structure factor $S(\mathbf{k})$ (\ref{skdef}) by
	\begin{equation}
		\lim_{N,V_F\rightarrow+\infty}\langle\mathcal{I}(\mathbf{k})\rangle=(2\pi)^d\rho\delta(\mathbf{k})+S(\mathbf{k}),
	\end{equation}
where $V_F$ is the volume of the fundamental cell and $\delta$ is the Dirac delta function \cite{To18a}. In simulations of many-body systems with finite $N$ under periodic boundary conditions, Eq. (\ref{scattering}) is used to compute $S(\mathbf{k})$ directly by averaging over configurations. The structure factor corresponding to the unit-step-function $g_2$ [Eq. (\ref{g2-us})] in dimensions $1,2,3$ are given by \cite{St05}
\begin{equation}
    S(k) = 
    \begin{cases}
    1- 2\rho\frac{\sin(k)}{k}, \quad d=1\\
    1 - 2\pi\rho\frac{J_1(k)}{k}, \quad d=2\\
    1 - 4\pi\rho\frac{\sin(k) - k\cos(k)}{k^3}, \quad d=3,
    \end{cases}
    \label{s_us}
\end{equation}
where $J_1(x)$ is the first-order Bessel function. Note that at  $\phi_c=1/2^d$, $S(k)\sim k^2$ as $k\rightarrow 0$, indicating that terminal-density systems are hyperuniform, if they are realizable.

\subsection{Hyperuniformity and nonhyperuniformity}
\label{hu}
A \textit{hyperuniform} point configuration in $d$-dimensional Euclidean space $\mathbb{R}^d$ is characterized by an anomalous suppression of large-scale density fluctuations relative to those in typical disordered systems, such as liquids and structural glasses \cite{To03a,To18a}. More precisely, a hyperuniform point pattern is one in which the structure factor $S(\mathbf{k}) = 1 + \rho\tilde{h}(\mathbf{k})$ tends to zero as the wave number $k = |\mathbf{k}|$ tends to zero \cite{To03a,To18a} i.e.,
\begin{equation}
    \lim_{|\mathbf{k}|\rightarrow 0} S(\mathbf{k}) = 0.
    \label{hyperuniformDef}
\end{equation}
This hyperuniformity condition implies the following direct-space sum rule:
\begin{equation}
    \rho \int_{\mathbb{R}^d}h(\mathbf{r})d\mathbf{r} = -1.
\end{equation}
An equivalent definition of hyperuniformity is based on the local number variance $\sigma^2(R)=\langle N(R)^2\rangle - \langle N(R)\rangle^2$ associated with the number $N(R)$ of points within a $d$-dimensional spherical observation window of radius $R$, where angular brackets denote an ensemble average. A point pattern in $\mathbb{R}^d$ is hyperuniform if its variance grows in the large-$R$ limit slower than $R^d$ \cite{To03a}. 

Consider systems that are characterized by a structure factor with a radial power-law form in the vicinity of the origin, i.e.,
\begin{equation}
    S(\mathbf{k})\sim |\mathbf{k}|^\alpha, \qquad \mbox{for } |\mathbf{k}| \rightarrow 0.
    \label{s_hu_smallk}
\end{equation}
For hyperuniform systems, the exponent $\alpha$ is positive ($\alpha > 0$) and its value determines three different large-$R$ scaling behaviors of the number variance \cite{To03a,Za09,To18a}: 
\begin{equation}
    \sigma^2(R)\sim \begin{cases}
          R^{d-1}, \quad \alpha > 1 \text{ (class I)} \\
          R^{d-1}\ln R, \quad \alpha=1 \text{ (class II)} \\
          R^{d-\alpha}, \quad 0<\alpha<1 \text{ (class III).} \\
     \end{cases}
\end{equation}
Torquato showed that for an equilibrium hyperuniform state under a pair potential whose structure factor is characterized by the power law (\ref{s_hu_smallk}), 
the potential $v(r)$ must be long-ranged in the sense that its volume integral is unbounded \cite{To18a}:
\begin{equation}
    v(r)\sim 
    \begin{cases}
    r^{-(d-\alpha)}, \quad d \ne \alpha\\
    -\log(r), \quad d = \alpha.
    \end{cases}
    \label{v-long-hu}
\end{equation}
Therefore, to stabilize a  classical hyperuniform system at positive $T$, which is thermodynamically incompressible \cite{To03a}, one must treat it as a system of ``like-charged'' particles immersed in a rigid ``background'' of equal and opposite ``charge'', i.e., the system must have overall charge neutrality \cite{To18a}. Such a rigid background contribution corresponds to a one-body potential, which is described in Sec. \ref{meth}.

By contrast, for any \textit{nonhyperuniform} system, the local variance has the following large-$R$ scaling behaviors \cite{To21c}:
\begin{equation}
    \sigma^2(R)\sim \begin{cases}
          R^{d}, \quad \alpha = 0 \text{ (typical nonhyperuniform)} \\
          R^{d-\alpha}, \quad \alpha<0 \text{ (antihyperuniform).} \\
     \end{cases}
\end{equation}
For a ``typical” nonhyperuniform system, $S(0)$ is bounded. In antihyperuniform systems \cite{To18a}, $S(0)$ is unbounded, i.e.,
\begin{equation}
    \lim_{|\mathbf{k}|\rightarrow 0} S(\mathbf{k}) = +\infty,
\end{equation}
and hence these systems are diametrically opposite to hyperuniform systems. Antihyperuniform systems include fractals, systems at thermal critical points (e.g., liquid-vapor and magnetic critical points) \cite{Wi65,Ka66,Fi67,Wi74,Bi92}, as well as certain substitution tilings \cite{Og19}.

\subsection{Order metric}
Scalar order/disorder metrics have been profitably used to quantify the degree of order in many-particle systems, including sphere packings \cite{To02a, To18b}. A particularly useful measure for our purposes is the order metric $\tau$ \cite{To15}, which is defined as  
\begin{equation}
\tau = \frac{1}{D^d}\int_{\mathbb{R}^d} h^2(\mathbf{r}) d\mathbf{r}
= \frac{1}{(2\pi)^d D^d} \int_{\mathbb{R}^d} \tilde{h}^2(\mathbf{k}) d\mathbf{k},
\label{tau}
\end{equation}
where $D$ is a characteristic ``microscopic'' length scale, taken to the the sphere diameter in this work.
This order metric measures deviations of two-particle statistics from that of the Poisson distribution. 
Since both positive and negative correlations contribute to the integral, due to the fact that $h(r)$ or $\tilde{h}(k)$ is squared, $\tau$ measures the degree of translational order across length scales. 
It clearly vanishes for the uncorrelated Poisson distribution, diverges for an infinite crystal and is a positive bounded number for correlated disordered systems without long-range order (i.e., Bragg peaks).

\subsection{Necessary conditions for realizability}
\label{nec}
Two well-known necessary conditions must be satisfied by any pair statistics \cite{Cr03, St05, Ko05, Uc06a}, namely, the nonnegativity of $g_2(r)$ and its corresponding $S(k)$:
\begin{equation}
    g_{2}(r) \geq 0, \quad \text{for all $r$},
    \label{g2_nonneg}
\end{equation}
\begin{equation}
    S(k) \geq 0, \quad \text{for all $k$}.
    \label{S_nonneg}
\end{equation}
A further necessary condition forces a lower bound on the variance associated with the number of particles within a $d$-dimensional spherical window of radius $R$ \cite{Ya61}:
\begin{equation}
    \sigma^2(R)\geq \theta(1-\theta),
\end{equation}
where $\theta$ is the fractional (non-integer) part of the average number of particle centers contained within the window $\langle N(R) \rangle= \rho v_1(R)$. The Yamada condition, in practice, is relevant in very low dimensions when (\ref{g2_nonneg}) and (\ref{S_nonneg}) are satisfied, and becomes less restrictive for higher dimensions \cite{To06b}. It turns out to pose no additional restriction on the realizability of all the target pair statistics considered in this study in any dimension \cite{To06b}.

\section{Inverse methodology}
\label{meth}
Here, we describe our methodology to determine (up to) pair interactions that yield canonical ensembles matching target pair statistics; see Ref. \citenum{To22} for more details.
This procedure enables one to determine equilibrium states that match both target $g_2(r)$ and $S(k)$ with unprecedented accuracy when they are realizable. 
Thus, it is especially suitable for probing the realizability of hypothesized functional form for the pair statistics, which heretofore has not been done.
The methodology uses a parameterized family of pointwise basis functions for the dimensionless pair potential
\begin{equation} 
v(r;{\bf a})= \varepsilon \sum_{j=1}^n f_j(r/\sigma;a_j), 
\label{basis} 
\end{equation} 
where $f_j(r/\sigma;a_j)$ is the $j$th basis function, $a_j$ is a vector of parameters (generally consisting of multiple parameters), ${\bf a}=(a_1,a_2,\ldots,a_n)$ is the ``supervector'' parameter, $\varepsilon$ sets the energy scale and $\sigma$ is a characteristic length scale. Both $\epsilon$ and $\sigma$ are taken to be unity.
The components of $a_j$ are of four types: dimensionless energy scales $\varepsilon_j$, dimensionless distance scales $\sigma_j$, dimensionless phases $\theta_j$, as well as dimensionless exponents $p_j$.

The initial form of $v(r;\mathbf{a})$ is informed by the small- and large-distance behaviors of the target pair statistics $g_{2,T}(r)$ and $S_T(k)$ as dictated by statistical-mechanical theory \cite{To18a}. Specifically, under mild conditions, the exact large-$r$ behavior of $v(r;{\bf a})$ is enforced in (\ref{basis}) to be the large-$r$ behavior of the targeted direct correlation function $c_T(r)$ via relation \cite{Stell77,Ha86}
\begin{equation}
c_T(r) \sim -\beta v(r;\mathbf{a}), \quad (|{\bf r}| \to +\infty),
\label{asymp}
\end{equation}
where $\beta =1/(k_B T)$ and $k_B$ is the Boltzmann constant, and the large-$|\bf r|$ behavior of $c({\bf r})$ can be extracted from the structure factor $S_T({\bf k})$ 
using the Fourier representation of the Ornstein-Zernike integral equation \cite{Or14}:
\begin{equation}
{\tilde c}({\bf k})= \frac{{\tilde h}({\bf k})}{S({\bf k})}.
\label{OZ}
\end{equation}
To estimate the initial small- and intermediate-$r$ behavior in (\ref{basis}), we simply use the hypernetted-chain (HNC) approximation \cite{Ha86}, i.e., 
 \begin{equation} 
 \beta v_{\text{HNC}}(r)= h_T(r)- c_T(r)- \ln[g_{2,T}(r)].
 \label{HNC}
\end{equation}

The basis functions are chosen so that they reasonably span all potential functions that could correspond to a targeted $g_{2,T}(r)$ for all $r$ or a targeted $S_T(k)$ for all $k$ under the constraint that the resulting potential function $v(r)$ satisfies the small-$r$ and large-$r$ behaviors dictated by statistical-mechanical theory described above. For our specific targets, $f_j(r;a_j)$ are chosen from the possible following general forms:
\begin{enumerate}
\item Hard core:
\begin{equation}
f_j(r;a_j)=
    \begin{cases}
    +\infty, \qquad &r \leq 1 \\
    0, \qquad &r > 1.
    \end{cases}
    \label{hard_core}
\end{equation}
This basis function is used for the unit-step-function $g_2$ and the RSA target, as their $g_{2,T}(r)$ exhibit a hard core for $r \leq 1$.
\item Exponential-damped oscillatory form:
\begin{equation}
    f_j(r;a_j)=\varepsilon_j\cos\left(\frac{r}{\sigma^{(1)}_j} + \theta_j \right)\exp\left[-\left(\frac{r-\sigma^{(2)}_j}{\sigma^{(3)}_j}\right)^M\right].
    \label{exponential_form}
\end{equation}
Importantly, the large-$r$ behavior of $v(r)\sim -\beta c(r)$ for the 1D unit-step-function $g_2$ for $\phi\rightarrow\phi_c^-$ (but $\phi\ne\phi_c$) is given by an exponential form \cite{To03a}; see Sec. \ref{unitstep}. In this work, for simplicity and efficiency, the exponent $M$ (\ref{exponential_form}) is restricted to be an integer that remains fixed during the optimization process as described below. 
\item Exponential-damped power-law form \cite{Yu55}:
\begin{equation}
    f_j(r;a_j)=
    \varepsilon_j\exp\left(-\kappa r \right)r^{-p_j}, \\
    \label{Yukawa_form}
\end{equation}
where $\kappa$ is the inverse screening length. This form is chosen, because it has been shown that the large-$r$ behavior of $v(r)\sim -\beta c(r)$ for the 2D and 3D unit-step-function $g_2$ in the limit for $\phi\rightarrow\phi_c^-$ (but $\phi\ne \phi_c$) is given by (\ref{Yukawa_form}) \cite{To03a}, with $p_j=1/2$ for $d=2$ and $p_j=1$ for $d=3$; see Sec. \ref{unitstep} for further details.
\item Power-law damped oscillatory form:
\begin{equation}
f_j(r;a_j)=\varepsilon_j\cos\left(\frac{r}{\sigma^{(1)}_j} + \theta_j \right)r^{-p_j}.
\label{power_form}
\end{equation}
\item Coulomb form (\ref{v-long-hu}), which is the large-$r$ behavior of $v(r)$ for the unit-step-function $g_2$ at the terminal packing fraction $\phi_c$ in all dimensions, given that the states are realizable; see Sec. \ref{hu}.
\end{enumerate}

Once the initial form of $v(r;\mathbf{a})$ is chosen, a nonlinear optimization procedure \cite{Liu89} is used to minimize an objective function $\Psi({\bf a})$ based on the distance between target and trial pair statistics in both direct and Fourier spaces:

\begin{equation}
\begin{split}
        \Psi(\mathbf{a}) &=\rho\int_{\mathbb{R}^d}
 w_{g_2}({\bf r})\left(g_{2,T}({\bf r})-g_{2}({\bf r};\mathbf{a})\right)^2
d{\bf r}\\ 
&+ \frac{1}{\rho(2\pi)^d}\int_{\mathbb{R}^d} w_{S}({\bf k})\left(\log(S_T({\bf k}))-\log(S(k;\mathbf{a}))\right)^2 d\mathbf{k},
\end{split}
\label{Psi}
\end{equation}

where $w_{g_2}({\bf r})$ and $w_{S}({\bf k})$ are weight functions, and $g_{2}({\bf r};\mathbf{a})$ and $S(\mathbf{k};\mathbf{a})$ correspond to an equilibrated $N$-particle system under $v(r;\mathbf{a})$ at a dimensionless temperature $k_BT/\varepsilon=1$, which can be obtained from Monte-Carlo (MC) (used here) or molecular dynamics simulations under periodic boundary conditions. The optimization procedure ends when $\Psi(\mathbf{a})$ is smaller than some small tolerance $\epsilon$. If this convergence criterion is not achieved, then a different set of basis functions is chosen and the optimization process is repeated. The additional basis functions are obtained via a fit of the difference of the potentials of mean forces between the simulated and target $g_2(r)$. 

For hyperuniform targets whose small-$k$ behavior is given by a power law $S(k)\sim k^\alpha$, $v(r;\mathbf{a})$ has the long-ranged asymptotic form given by $v(r)\sim 1/r^{d-\alpha}$, which can be regarded as a generalized Coulombic interaction of ``like-charged'' particles \cite{To18a}. Thus, one requires a neutralizing background one-body potential to maintain stability \cite{To18a, Ha73, Ga79, Dy62a}. Importantly, to perform the MC simulations, the total potential energy involving the long-ranged one- and two-body potentials is efficiently evaluated using the Ewald summation technique \cite{Ew21}.

In our implementation of the inverse procedure, we used $w_{g_2}(r)=\exp(-r^2/16)$, $w_{S}(k)=\exp(-k^2/4)$, $\epsilon=5\times 10^{-4}$, and the system sizes in MC simulations were $N=400, 500, 1000$ for $d=1,2,3$, respectively. However, if a hyperuniform target is found to be numerically realizable for the aforementioned small system size, we proceed to verify its realizability for larger system sizes, i.e., $N=10,000$ for $d=1, 2$ and $N=20,000$ for $d=3$.
If convergence is not achieved within 5 iterations of re-selection of basis functions, the target is deemed to be potentially not realizable. 

\section{Iso-$g_2$ process for the unit-step-function $g_2$}
\label{unitstep}

In this section, we present the realizable density ranges for the optimized effective interactions for the unit-step-function $g_2$ for $d=1,2,3$. Taylor expansions of Eq. (\ref{s_us}) at the point $k=0$ reveal that $S(k)$ for the unit-step-function $g_2$ are analytic at the origin, implying that the power series only admits even powers of $k$ \cite{To18a}. Thus, the effective potential must have exponential or superexponential decay at $\phi<\phi_c$, and a Coulomb form (\ref{v-long-hu}) at $\phi=\phi_c$ \cite{To18a, To22}.
Indeed, Torquato and Stillinger showed that at $\phi=\phi_c$, the target direct correlation function for $r \gg 1$ is given by \cite{To03a, Note5} 
\footnotetext[5]{Equations (\ref{ct_phic}), (\ref{cT_notPhic_alld}) and (\ref{kappa}) in the present work are Eqs. (118), (120) and (114) in Ref. \citenum{To03a}, respectively, where $\xi=\kappa^{-1}$. However, Eqs. (114) and (120) in Ref. \citenum{To03a} contain trivial errors in constants due to a typo introduced in Eq. (111). The correct form of (111) in Ref. \citenum{To03a} is 
\begin{equation}
    \tilde{c}(k)=\frac{-v_1(D)}{(1-\frac{\phi}{\phi_c})+\frac{1}{2(d+2)}(kD)^2+O((kD)^4)}.
\end{equation}
Equation (119) in Ref. \citenum{To03a} is also incorrect due to this typo. The correct form of (119) is
\begin{equation}
    c(r)= 
    \begin{cases}
        6\left(\frac{\xi}{D}\right)\exp(-r/\xi), \quad d=1\\
        4 \ln\left(\frac{r}{D}\right)\exp(-r/\xi), \quad d=2\\
        - \frac{2(d+2)}{d(d-2)} \left(\frac{r}{D}\right)^{-(d-2)}\exp(-r/\xi), \quad d\geq 3.
    \end{cases}
    \label{ct_large_xi}
\end{equation}
We note that Eqs. (111), (114), (119), and (120) are the only equations affected by the typo. These errors do not affect the main conclusions of Ref. \citenum{To03a}.
}
\begin{equation}
    c_T(r)\sim
    \begin{cases}
        6r, \quad d=1\\
        4 \ln(r), \quad d=2\\
        - \frac{2(d+2)}{d(d-2)} r^{-(d-2)}, \quad d\geq 3.
    \end{cases}
    \label{ct_phic}
\end{equation}
Furthermore, in the limit $\phi\rightarrow\phi_c^-$ (but $\phi\ne\phi_c$), the form of $c_T(r)$ at large $r$ is given by an exponential form for $d=1$, and by exponential-damped power-law forms for $d\geq 2$, i.e. \cite{To03a}, 
\begin{equation}
    c_T(r)\sim -\frac{(d+2)\sqrt{2\pi}}{\sqrt{2^d}\Gamma(1+d/2)}\kappa(\phi)^{\frac{d-3}{2}}r^{-\frac{d-1}{2}}\exp[-\kappa(\phi) r],
    \label{cT_notPhic_alld}
\end{equation}
where $\kappa(\phi)$ is the inverse screening length that depends on the packing fraction \cite{To03a}: 
\begin{equation}
    \kappa(\phi) = \sqrt{2(d+2)(1-\frac{\phi}{\phi_c})}.
    \label{kappa}
\end{equation}
Specifically, in the first three dimensions, one has
\begin{equation}
    c_T(r)\sim 
    \begin{cases}
        -6\kappa^{-1}(\phi)\exp[-\kappa(\phi) r], \quad d=1\\
        -2\sqrt{\frac{2\pi}{\kappa(\phi) r}}\exp[-\kappa(\phi) r], \quad d=2\\
        -  \frac{10}{3r} \exp[-\kappa(\phi) r], \quad d= 3;
    \end{cases}
    \label{cT_notPhic}
\end{equation}
note that the case $d=3$ is a Yukawa form.
We numerically fitted $-c_T(r)/\beta\sim v(r)$ and found that the large-$r$ behaviors of $v(r;\mathbf{a})$ are indeed well-described by (\ref{ct_phic}) for $\phi=\phi_c$ and by (\ref{cT_notPhic}) for $7/8\leq\phi/\phi_c<1$. Thus, we use these forms as the longest-ranged basis functions for $v(r;\mathbf{a})$.

\subsection{1D systems}
\begin{figure*}[htp]
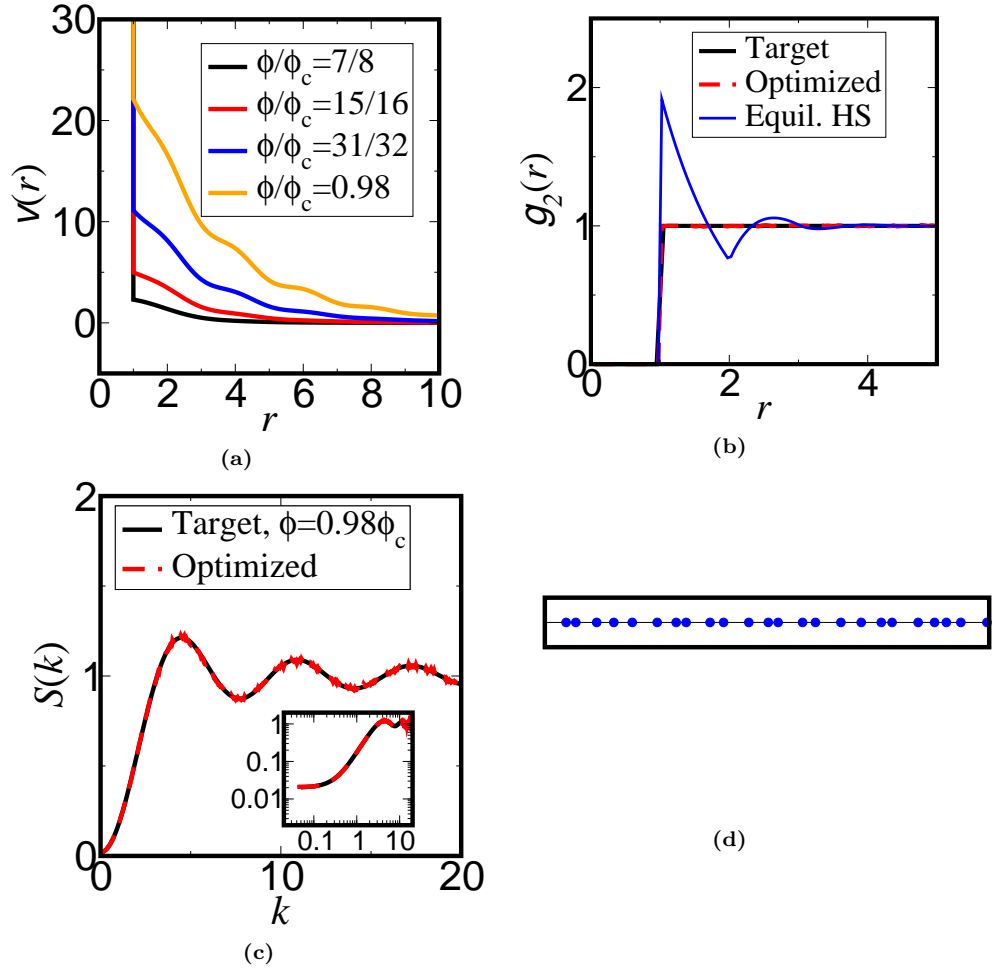

  \centering
  \begin{subfigure}{6cm}
    \includegraphics[width=60mm]{Fig2a.eps}
    \caption{}
    \label{2a}
  \end{subfigure}
  \qquad
  \begin{subfigure}{5.6cm}
    \includegraphics[width=56mm]{Fig2b.eps}
    \caption{}
    \label{2b}  
  \end{subfigure}
  
  \begin{subfigure}{6cm}
    \includegraphics[width=60mm]{Fig2c.eps}
    \caption{}
    \label{2c}  
  \end{subfigure}
  \qquad
  \begin{subfigure}{5cm}
    \includegraphics[trim={0 -9cm 0 0},clip,width=60mm]{Fig2d.eps}
    \caption{}
    \label{2d}  
  \end{subfigure}
\captionsetup{subrefformat=parens}
  \caption{\subref{2a} The effective potentials for the 1D unit-step-function $g_2$ (\ref{g2-us}).
  \subref{2b} Target and optimized $g_2$ at the maximum realizable packing fraction $\phi_m=0.49=0.98\phi_c$. Here we find that the $L_2$ function (\ref{g2-norm}) is $D_{g_2}=6\times 10^{-4}$. 
  The blue curve shows $g_2(r)$ for the pure equilibrium hard-rod fluid at $\phi_m$.
  \subref{2c} Target and optimized $S(k)$ at $\phi_m$. Here we find that the $L_2$ function (\ref{S-norm}) is $D_S=0.0010$. The $L_2$-norm error (\ref{L2}) is $\mathcal{E}=0.040$. The inset shows on a log scale the small-$k$ behaviors. 
  \subref{2d} A 25-particle portion of a 400-particle configuration of the system with unit-step-function $g_2$ at $\phi_m$.}
  \label{fig:1DUnitStep}
\end{figure*}

The long-ranged part of effective potential for the unit-step-function $g_2$ in one dimension is has exponential decay \cite{To03a}, i.e.,
\begin{equation}
    v(r;\phi)\sim
    \exp[-\kappa(\phi) r],
    \quad r\rightarrow+\infty.
\end{equation}
We find that the maximum density at which the 1D unit-step-function $g_2$ is numerically realizable is $\phi_m=0.49=0.98\phi_c$. The functional form of these potentials is given by
\begin{widetext}
\begin{equation}
    v(r;\phi)=
    \begin{cases}
    +\infty, \quad r \leq 1 \\  \varepsilon_1\exp[- \kappa(\phi) r] +  \sum_{j=2}^4 \varepsilon_j\exp\left(-\frac{r}{\sigma_j^{(1)}}\right)\cos\left(\frac{r}{\sigma_j^{(2)}}+\theta_j\right)   + \\
    \sum_{j=5}^6 \varepsilon_j\exp\left[-\left(\frac{r}{\sigma_j^{(1)}}\right)^2\right]\cos\left(\frac{r}{\sigma_j^{(2)}}+\theta_j\right),
     \quad r> 1.
    \end{cases}
    \label{v_unitstep_1D}
\end{equation}
\end{widetext}
Table \ref{1D_unitStep_params} lists the values of the optimized parameters for $\phi/\phi_c=7/8, 15/16, 31/32$ and 0.98, where it is shown that the larger $\phi$ is, the more oscillatory terms must be included. Figure \ref{fig:1DUnitStep}(\subref{2a}) shows the density-dependent effective potential at the aforementioned packing fractions, which clearly shows that for $\phi/\phi_c \geq 31/32$, $v(r;\phi)$ exhibits significant oscillations (or ``wiggles'') to negate the oscillations in $g_2(r)$. The amplitude of the sum of the oscillatory terms increases with $\phi$. 
As we will show in Sec. \ref{2d-sys} and \ref{3d}, such oscillatory behavior is insignificant in $v(r,\phi)$ for 2D and 3D unit-step-function $g_2$ due to the decorrelation principle in higher dimensions \cite{To18a}. 
Figures \ref{fig:1DUnitStep}(\subref{2b}) and \ref{fig:1DUnitStep}(\subref{2c}) show the target and optimized pair statistics at the maximum realizable packing fraction $\phi_m=0.49=0.98\phi_c$. 
We find that the $L_2$ functions [Eqs. (\ref{g2-norm})--(\ref{S-norm})] are $D_{g_2}=6\times 10^{-4}$, $D_S=0.0010$ and the $L_2$-norm error (\ref{L2}) is $\mathcal{E}=0.040$. These errors are an order of magnitude smaller than typical errors obtained via IHNCI for realizable targets \cite{To22}. 
Figure \ref{fig:1DUnitStep}(\subref{2b}) also shows $g_2(r)$ for the pure equilibrium hard-rod system at $\phi_m=0.49$, which exhibits significant positive and negative correlations compared to the unit-step-function $g_2$. 
This demonstrates that the unit-step-function-$g_2$ system is much more disordered than the pure equilibrium hard-rod system, and that the effective potential (\ref{v_unitstep_1D}) significantly suppresses the correlations beyond the hard core, as will be quantitatively analyzed in Sec. \ref{kappa_tau} via the order metric $\tau$ (\ref{tau}).
Figure \ref{fig:1DUnitStep}(\subref{2d}) shows a 25-particle portion of a 400-configuration at $\phi_m$. 

\begin{table}
\caption{Optimized parameters of the effective pair potential for the 1D unit-step-function $g_2$.}
\begin{tabular}{ ||c|c|c|c|c||c|c|c|| } 
 \hline
 $\phi/\phi_c$ & 7/8 & 15/16 & 31/32 & 0.98 & $\phi/\phi_c$ & 31/32 & 0.98 \\ 
 \hline
 $\varepsilon_1$ & 6.47 & 10.12 & 11.18 & 29.52 & $\varepsilon_4$ & -4.098 & -1.828 \\
 $\kappa^{-1}(\phi)$ & 1.165 & 1.641 & 2.440 & 2.704 & $\sigma_4^{(1)}$ & 0.5710 & 3.152 \\
 $\varepsilon_2$ & 2.809 & 0.7187 & -6.772 & -3.056 & $\sigma_4^{(2)}$ & 0.2687 & 0.3254 \\
 $\sigma_2^{(1)}$ & 0.7674 & 1.680 & 1.094 & 0.6015 & $\theta_4$ & 4.000 & 3.077 \\
 $\sigma_2^{(2)}$ & -0.5470 & -0.3393 & 1.375 & -0.2195 & $\varepsilon_5$ & 0.1934 & 0.04138 \\
 $\theta_2$ & 4.059 & 0.1577 & 2.050 & 3.332 & $\sigma_5^{(1)}$ & 60.23 & 21.99 \\
 $\varepsilon_3$ & \dots & 10.74 & -1.534 & -5.844 & $\sigma_5^{(2)}$ & -425.8 & 2.073 \\
 $\sigma_3^{(1)}$ & \dots & 0.5624 & 2.211 & 1.587 & $\theta_5$ & 1.781 & 4.089 \\
 $\sigma_3^{(2)}$ & \dots & 0.7691 & 0.3275 & 1.420 & $\varepsilon_6$ & 3.494 & 0.3899 \\
 $\theta_3$ & \dots & 3.317 & 2.960 & 2.902 & $\sigma_6^{(1)}$ & 4.280 & 6.914 \\
  &  &  &  & & $\sigma_6^{(2)}$ & 19.36 & 1.629 \\
 &  &  &  & & $\theta_6$ & 0.8264 & 4.124 \\
 \hline
\end{tabular}
\label{1D_unitStep_params}
\end{table}

\begin{figure*}[htp]
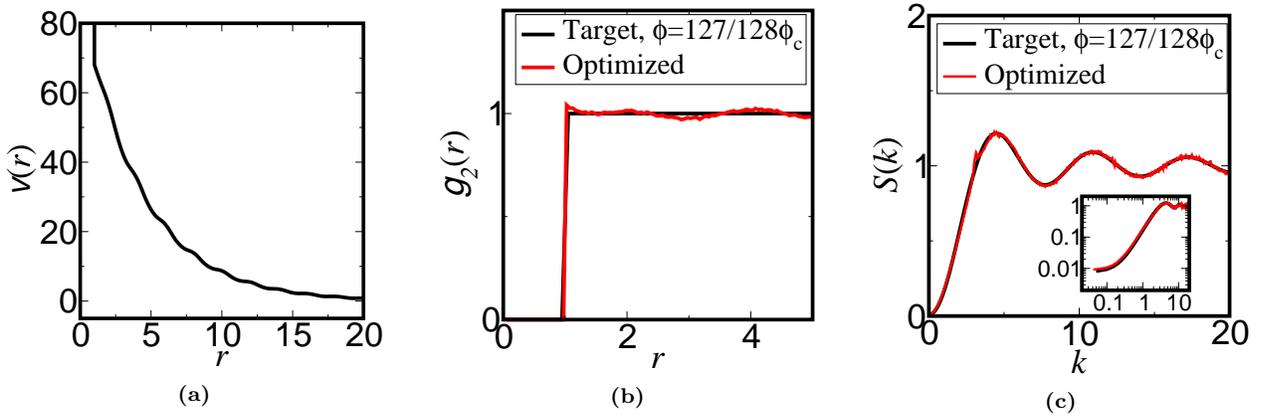

  \centering
  \begin{subfigure}{5cm}
    \includegraphics[width=50mm]{Fig3a.eps}
    \caption{}
    \label{3a}
  \end{subfigure}
  \qquad
  \begin{subfigure}{5cm}
    \includegraphics[width=50mm]{Fig3b.eps}
    \caption{}
    \label{3b}  
  \end{subfigure}
  \qquad
  \begin{subfigure}{5cm}
    \includegraphics[width=50mm]{Fig3c.eps}
    \caption{}
    \label{3c}  
  \end{subfigure}
\captionsetup{subrefformat=parens}
  \caption{\subref{3a} The optimized trial potential (\ref{v_unitstep_1D}) for the 1D unit-step-function $g_2$  (\ref{g2-us}) at $\phi=127/256=127/128\phi_c$. \subref{3b} Target and optimized $g_2(r)$. \subref{3c} Target and optimized $S(k)$. The inset shows on a log scale the small-$k$ behaviors.}
  \label{fig:1DUnitStep127-128}
\end{figure*}

Figure \ref{fig:1DUnitStep127-128}(\subref{3a}) shows the optimized potential of the form (\ref{v_unitstep_1D}) at packing fraction $\phi=127/128\phi_c>\phi_m$, obtained after 5 iterations of re-selecting the basis functions. 
Figures \ref{fig:1DUnitStep127-128}(\subref{3b}) and \ref{fig:1DUnitStep127-128}(\subref{3c}) show the corresponding optimized pair statistics. It is clear that the target pair statistics are not realized by the simulated many-body system. The optimized $g_2(r)$ oscillates around unity for $r>1$, and contains a peak at $r=1$. Thus, the system exhibits clustering of particles despite the highly repulsive nature of the potential. The optimized $S(k)$ contains a small peak at $k\sim \pi$, and the $S(k)$ values at small $k$ are higher than those of the target $S_T(k)$. 

To test whether the non-convergence of the trial potential is due to the constraints of the forms of the basis functions, we included two additional exponential-damped oscillatory basis functions (\ref{exponential_form}) to (\ref{v_unitstep_1D}) and repeated the optimization procedure. 
We observed that convergence was still not reached,  
which suggests that the target is potentially indeed non-realizable.  

We also applied the IHNCI procedure \cite{Le85,He18}, which was the most accurate inverse algorithm prior to the development of our methodology \cite{To22}, on the 1D unit-step-function $g_2$ target at $\phi=127/256=127/128\phi_c$.
We found that the trial $S(k)$ at small $k$ is larger than the target, and the trial $v(r)$ at all $r$ increased in each iteration, due to the fact that the algorithm attempts to match the small values of $S_T(k)$ around the origin. In about 100 iterations, the trial potential became so repulsive that the system crystallized, leading to a trial $g_2(r)$ that is drastically different from $g_{2,T}(r)$. Therefore, IHNCI is definitely not suited to treat the realizability problem in this case.

\subsection{2D systems}
\label{2d-sys}
\begin{figure*}[htp]
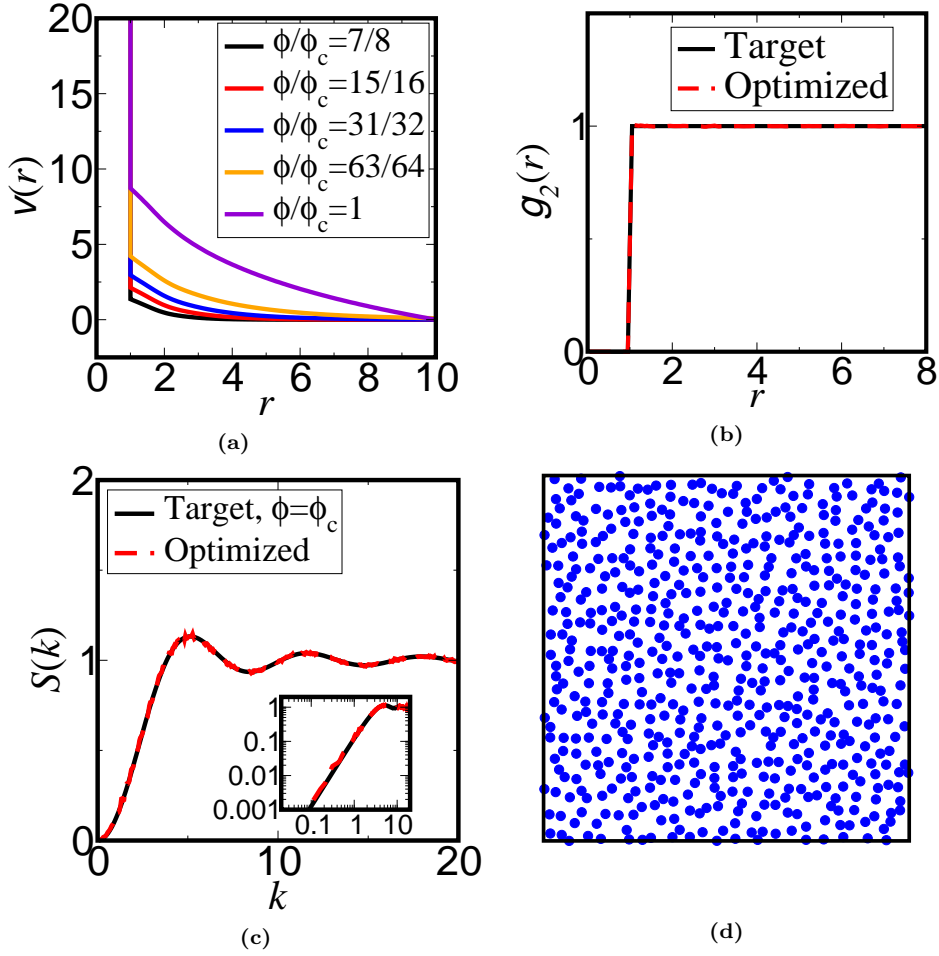

  \centering
  \begin{subfigure}{6cm}
    \includegraphics[width=60mm]{Fig4a.eps}
    \caption{}
    \label{4a}
  \end{subfigure}
  \qquad
  \begin{subfigure}{5.6cm}
    \includegraphics[width=56mm]{Fig4b.eps}
    \caption{}
    \label{4b}  
  \end{subfigure}
  
  \begin{subfigure}{6cm}
    \includegraphics[width=60mm]{Fig4c.eps}
    \caption{}
    \label{4c}  
  \end{subfigure}
  \qquad
  \begin{subfigure}{5cm}
    \includegraphics[trim={0 -4cm 0 0},clip,width=50mm]{Fig4d.eps}
    \caption{}
    \label{4d}  
  \end{subfigure}
\captionsetup{subrefformat=parens}
    \caption{\subref{4a} The effective potentials for the 2D unit-step-function $g_2$ (\ref{g2-us}). The potential for $\phi_c=1/4$ is shifted such that $v(10;\phi_c)=0$. \subref{4b} Target and optimized $g_2(r)$ at $\phi_c$.  Here we find that the $L_2$ function (\ref{g2-norm}) is $D_{g_2}=6\times 10^{-4}$. \subref{4c} Target and optimized $S(k)$ at $\phi_c$. Here we find that the $L_2$ function (\ref{S-norm}) is $D_S=0.0022$. The $L_2$-norm error (\ref{L2}) is $\mathcal{E}=0.053$. The inset shows on a log scale the small-$k$ behaviors. \subref{4d} A 600-particle configuration of the 2D system with unit-step-function $g_2$ at $\phi_c$.}
  \label{fig:2Dunitstep}
\end{figure*}

The long-ranged part of effective potential for the unit-step-function $g_2$ in two dimensions is determined to be the exponentially-screened inverse power-law $r^{-1/2}$\cite{To03a} for $\phi<\phi_c$, i.e.,
\begin{equation}
    v(r;\phi)\sim
    \exp[-\kappa(\phi) r]r^{-1/2},
    \quad r\rightarrow+\infty.
\end{equation}
We find that the 2D unit-step-function $g_2$ is numerically realizable up to the terminal density $\phi_c=1/4$. Figure \ref{fig:2Dunitstep}(\subref{4a}) shows the density-dependent effective potential $v(r,\phi)$, whose functional form is given by
\begin{equation}
    v(r;\phi)=
    \begin{cases}
    +\infty, \quad r \leq 1 \\
    \varepsilon_1 r^{-1/2}\exp[-\kappa(\phi) r] 
    \\+ \varepsilon_2\exp\left(-r/\sigma_2^{(1)}\right)\cos(r/\sigma_2^{(2)}+\theta_2), \quad r> 1.
    \end{cases}
    \label{v_unitstep_2D}
\end{equation}
Table \ref{2D_unitStep_params} lists the optimized values of the parameters. The potential becomes increasingly long-ranged with increasing $\phi$, and the characteristic length scale $\kappa^{-1}(\phi)$ diverges to infinity as $\phi\rightarrow \phi_c^-$. At $\phi=\phi_c$, the long-ranged part of the potential becomes the 2D Coulomb form:
\begin{equation}
    v(r, \phi_c)=
    \begin{cases}
    +\infty, \quad r\leq 1 \\
    -\varepsilon_1\log(r)+\\ \varepsilon_2\exp\left(-r/\sigma_2^{(1)}\right)\cos(r/\sigma_2^{(2)}+\theta_2), \quad r>1.
    \end{cases}
\end{equation}
As shown in Fig. \ref{fig:2Dunitstep}(\subref{4b}), the equilibrium state at $\phi_c$ accurately yields the unit-step-function $g_2$.
Figure \ref{fig:2Dunitstep}(\subref{4c}) shows that the optimized system is hyperuniform, and its structure factor matches $S_T(k)$. 
As in the 1D case, the precision of our methodology is evident from the small values of the $L_2$ functions [Eqs. (\ref{g2-norm})--(\ref{S-norm})] and the $L_2$-norm error (\ref{L2}), given by $D_{g_2}=6\times 10^{-4}$, $D_S=0.0022$ and $\mathcal{E}=0.053$.
Figure \ref{fig:2Dunitstep}(\subref{4d}) shows a snapshot of the optimized system at $\phi_c$. The particles are well-separated and do not form large clusters.

\begin{center}
\begin{table}
\caption{Optimized parameters of the effective pair potential for the 2D unit-step-function $g_2$.}
\begin{tabular}{ ||c|c|c|c|c|c|c|| } 
 \hline
 $\phi/\phi_c$ & $\varepsilon_1$ & $\kappa^{-1}(\phi)$ & $\varepsilon_2$ &  $\sigma_2^{(1)}$ & $\sigma_2^{(2)}$ & $\theta_2$\\ 
 \hline
 7/8 & 5.254 & 0.9684 & 54.40 & 0.2182 & -0.3018 & 0.6078  \\
 15/16 & 5.430 & 1.455 & 595.7 & 0.1612 & -0.4543 & 0.1322 \\
 31/32 & 5.977 & 2.093 & 25.06 & 0.2881 & -2.857 & 3.502 \\
 63/64 & 6.665 & 3.425 & -20.50 & 0.3172 & 2.854 & -0.4685  \\
 1 & 4. & $+\infty$ & 6.774 & 0.4250 & 0.4834 & 1.776 \\ 
 \hline
\end{tabular}
\label{2D_unitStep_params}
\end{table}
\end{center}

\subsection{3D systems}
\label{3d}
As in the case of the 1D and 2D unit-step-function $g_2$, the long-ranged part of $v(r;\mathbf{a})$ for the 3D unit-step-function $g_2$ is found to be the Yukawa form \cite{To03a}. The target pair statistics is realizable up to $\phi_c=1/8$, and the optimized density-dependent pair potential is given by
\begin{equation}
    v(r;\phi)=
    \begin{cases}
    +\infty, \quad r \leq 1 \\
    \varepsilon_1 \exp[-\kappa(\phi) r]/r \\+ \varepsilon_2 \exp\left(-r/\sigma_2^{(1)}\right)\cos(\sigma_2^{(2)} + \theta_2), \quad r> 1,
    \end{cases}
    \label{v_unitstep_3D}
\end{equation}
where $\kappa(\phi_c) = 0$, giving a long-ranged potential with a Coulomb tail, i.e., $v(r;\phi_c)\sim 1/r$. 

\begin{figure*}[htp]
  \centering
  \begin{subfigure}{6cm}
    \includegraphics[width=60mm]{Fig5a.eps}
    \caption{}
    \label{5a}
  \end{subfigure}
  \qquad
  \begin{subfigure}{5.6cm}
    \includegraphics[width=56mm]{Fig5b.eps}
    \caption{}
    \label{5b}  
  \end{subfigure}
  
  \begin{subfigure}{6cm}
    \includegraphics[width=60mm]{Fig5c.eps}
    \caption{}
    \label{5c}  
  \end{subfigure}
  \qquad
  \begin{subfigure}{5.5cm}
    \includegraphics[width=55mm]{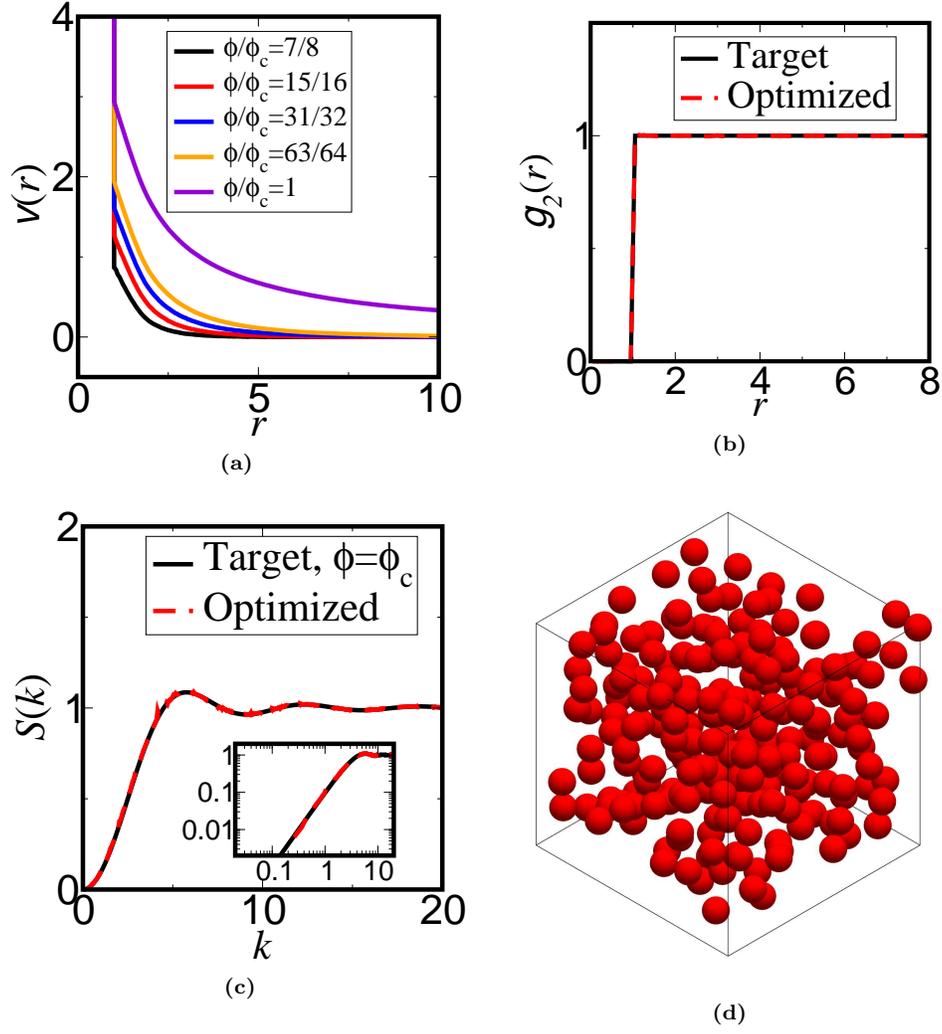}
    \caption{}
    \label{5d}  
  \end{subfigure}
  \captionsetup{subrefformat=parens}
  \caption{\subref{5a} The effective potentials for the 3D unit-step-function $g_2$ (\ref{g2-us}). \subref{5b} Target and optimized $g_2(r)$ at the terminal packing fraction $\phi_c=1/8$. Here we find that the $L_2$ function (\ref{g2-norm}) is $D_{g_2}=4\times 10^{-4}$. \subref{5c} Target and optimized $S(k)$ at $\phi_c$. Here we find that the $L_2$ function (\ref{S-norm}) is $D_S=0.0016$. The $L_2$-norm error (\ref{L2}) is $\mathcal{E}=0.045$. The inset shows on a log scale the small-$k$ behaviors. \subref{5d} A 216-particle configuration of the 3D system with unit-step-function $g_2$ at $\phi_c$.}
  \label{fig:3DUnitStep}
\end{figure*}

Figure \ref{fig:3DUnitStep}(\subref{5a}) shows the density-dependent effective potential of the unit-step-function $g_2$ for packing fractions $\phi$ in the vicinity of $\phi_c$. As $\phi$ increases, $v(r;\phi)$ becomes increasingly long-ranged. Table \ref{3D_unitStep_params} lists the optimized parameters in $v(r;\phi)$, from which it is obvious that the screening length of the potential $\kappa^{-1}(\phi)$ increases with $\phi$. Figures \ref{fig:3DUnitStep}(\subref{5b}) and \ref{fig:3DUnitStep}(\subref{5c}) show that the pair statistics at $\phi_c$ is excellently produced by the equilibrium system under the effective potential, 
as manifested by the small $L_2$ functions [Eqs. (\ref{g2-norm})--(\ref{S-norm})] $D_{g_2}=4\times 10^{-4}$ and $D_S=0.0016$, as well as the small $L_2$-norm error (\ref{L2}) $\mathcal{E}=0.045$.
Figure \ref{fig:3DUnitStep}(\subref{5d}) shows a snapshot of the system with unit-step-function $g_2$ at $\phi_c$.

\subsection{Screening lengths and order metrics across dimensions}
\label{kappa_tau}

\begin{figure*}[htp]
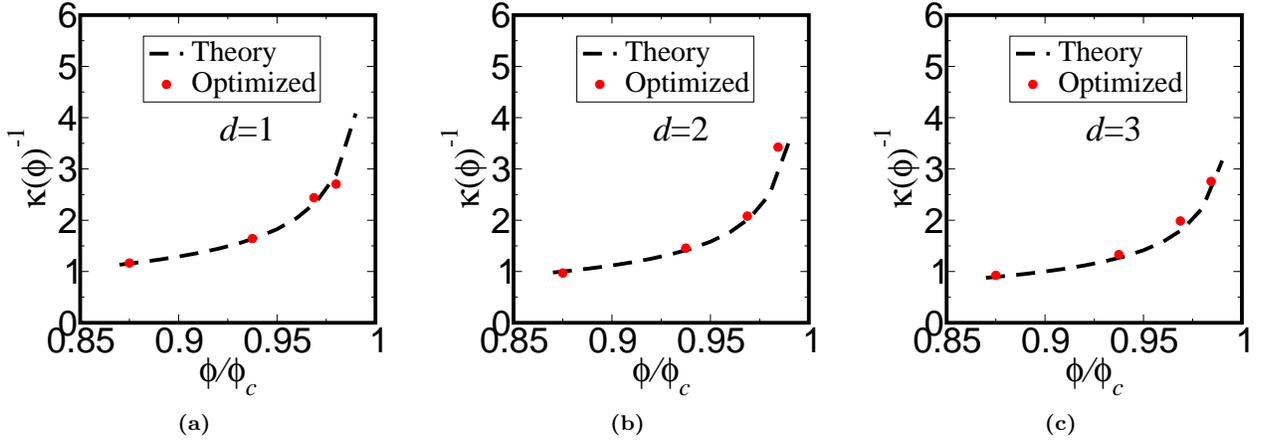

  \centering
  \begin{subfigure}{5cm}
  \includegraphics[width=50mm]{Fig6a.eps}
  \caption{}
  \label{6a}
  \end{subfigure}
  \qquad
  \begin{subfigure}{5cm}
  \includegraphics[width=50mm]{Fig6b.eps}
  \caption{}
  \label{6b}
  \end{subfigure}
  \qquad
  \begin{subfigure}{5cm}
  \includegraphics[width=50mm]{Fig6c.eps}
  \caption{}
  \label{6c}
  \end{subfigure}
  \captionsetup{subrefformat=parens}
  \caption{The potential screening length $\kappa^{-1}(\phi)$ as a function of the relative packing fraction $\phi/\phi_c$ for \subref{6a} 1D, \subref{6b} 2D, and \subref{6c} 3D unit-step-function $g_2$, where $\phi_c=1/2^d$ is the terminal packing fraction. The dashed curves show the theoretical relation (\ref{kappa}) and the filled circles are values obtained from the optimized effective potentials (\ref{v_unitstep_1D}), (\ref{v_unitstep_2D}) and (\ref{v_unitstep_3D}).}
  \label{fig:xi}
\end{figure*}

To study the dependence of the screening length $\kappa^{-1}(\phi)$ on the packing fraction and the dimensionality, we plot in Fig. \ref{fig:xi} the optimized $\kappa^{-1}(\phi)$ against $\phi/\phi_c$ for $d=1,2$ and 3, as well as the theoretical relation (\ref{kappa}) shown as dashed curves. 
In all dimensions, the theoretical and optimized $\kappa^{-1}(\phi)$ agree well, revealing the applicability of Eqs. (\ref{cT_notPhic}) and (\ref{kappa}) for $\phi$ near but below $\phi_c$. 
Nevertheless, note that the discrepancy between theoretical and optimized $\kappa^{-1}(\phi)$ is larger (on the order of 10\%) for $\phi/\phi_c\geq 31/32$ in all dimensions, due to the decreased sensitivity of pair statistics on pair potentials for higher-density states \cite{Wa20}. 
Importantly, for the same value of $\phi/\phi_c$, $\kappa^{-1}(\phi)$ decreases with $d$, reflecting the decorrelation principle for higher dimensions \cite{To18a}. 

To quantitatively characterize the suppression of correlations in the  unit-step-function $g_2$ relative to the corresponding equilibrium system with pure HS interactions at the same $\phi$, we compute their order metrics $\tau$ (\ref{tau}) for $d=1,2,3$ at the maximum realizable packing fractions for the unit-step-function $g_2$.
Equation (\ref{tau}) yields immediately that $\tau = v_1(1)$ for the unit-step-function-$g_2$ systems. 
The explicit expression of $S(k)$ for the 1D pure equilibrium HS system is given in Ref. \citenum{To21c}, which yields $\tau = 2.43$ at $\phi_m=0.49=0.98\phi_c$.
Note that $\tau$ for the unit-step-function $g_2$ at $\phi_m$ is only 41\% of that for the equilibrium HS system (Table \ref{unitStep_tau}), indicating that the effective potential (\ref{v_unitstep_1D}) achieves significant suppression of both positive and negative correlations that would otherwise be present in the pure equilibrium HS fluid.

The equilibrium pair statistics for the 2D and 3D pure equilibrium HS systems are not known exactly \cite{To02a}. 
Thus, we numerically computed $\tau$ from the pair statistics obtained via MC simulations. 
Table \ref{unitStep_tau} shows the $\tau$ order metrics (\ref{tau}) for the unit-step-function-$g_2$ systems and the pure equilibrium HS packings.
We observe that the discrepancy in the $\tau$ values for the two systems decreases as $d$ increases, which again reflects the decorrelation principle for higher dimensions \cite{To18a}. 

\begin{table}
\caption{Order metric (\ref{tau}) for unit-step-function-$g_2$ systems at their highest realizable packing fractions ($\tau_{\text{US}}$) and for pure equilibrium HS packings at the same $\phi$ ($\tau_{\text{HS}}$).}
\begin{tabular}{||c|c|c|c|c||}
\hline
$d$ & $\phi$ & $\tau_{\text{US}}$ & $\tau_{\text{HS}}$ & $\tau_{\text{US}}/\tau_{\text{HS}}$ \\
\hline
1 & $0.98\phi_c=0.49$ & 1 & 2.43 & 41\% \\
2 & $\phi_c=1/4$ & $\pi \approx 3.14$ & 3.72 & 85\% \\
3 & $\phi_c=1/8$ & $4\pi/3\approx 4.19$ & 4.57 & 92\% \\
\hline
\end{tabular}
\label{unitStep_tau}
\end{table}

\begin{center}
\begin{table}
\caption{Optimized parameters of the effective pair potential for the 3D unit-step-function $g_2$.}
\begin{tabular}{ ||c|c|c|c|c|c|c|| } 
 \hline
 $\phi/\phi_c$ & $\varepsilon_1$ & $\kappa^{-1}(\phi)$ & $\varepsilon_2$ & $\sigma_2^{(1)}$ & $\sigma_2^{(2)}$ & $\theta_2$ \\ 
 \hline
 $7/8$ & 3.227 & 0.926 & 13.92 & 0.2615 & -0.3368 & 5.368 \\ 
 $15/16$ & 3.262 & 1.352 & 15.84 & 0.2528 & -0.2850 & 0.06881 \\
 $31/32$ & 3.166 & 1.985 & -13.96 & 0.2703 & -0.3203 & 2.610 \\
 $63/64$ & 3.295 & 2.751 & -25.47 & 0.2342 & -0.3079 & 3.102 \\
 $1$ & 3.378 & $+\infty$ & -450.3 & 0.1521 & -0.3720 & 3.502 \\
 \hline
\end{tabular}
\label{3D_unitStep_params}
\end{table}
\end{center}

\section{3D ghost RSA target}
\label{ghost}
In this section, we compare the effective potential for the unit-step-function $g_2$ with that for a closely related model known as the ``ghost'' RSA process \cite{To06a, To06b}. It is a special case of a generalization of the standard RSA process \cite{Re63, wi66, To02a} and is the only known model for which all $n$-particle correlation functions are exactly solvable \cite{To06a}. 
In the ghost RSA process, spherical ``test'' particles of unit diameter are added continually to $\mathbb{R}^d$ during time $t\geq 0$ according to a translationally invariant Poisson process of density $\eta$ per unit time, i.e., $\eta$ is the number of points per unit volume and time, here taken to be unity without loss of generalization. A test sphere centered at position $\mathbf{r}$ at time $t$ is retained if and only if it does not overlap with any test sphere added in the time interval $[0,t)$ \cite{To06a}. The saturation packing fraction for the ghost RSA process is $\phi(t=+\infty)=1/2^d$, identical to the terminal density $\phi_c$ of the iso-$g_2$ process for the unit-step-function $g_2$ \cite{To06a}. 
For finite $d$, the ghost RSA model is nonhyperuniform at $\phi_c$, and the corresponding $g_2(r)$ possesses a small ``bump'' at $r=1$ and is identically unity for $r\geq 2$. The existence of such a bump at saturation density might suggest that the unit-step-function $g_2$ is unrealizable at $\phi_c$ for small $d$. Here, we study the pair statistics of the 3D ghost RSA process at saturation ($\phi_c=1/8$), whose $g_{2, T}(r)$ is given by \cite{To06a}
\begin{equation}
    g_{2,T}(r)=\frac{2\Theta(r-1)}{2-[(1-\frac{4r}{3}+\frac{r^3}{16})\Theta(2-r)]},
    \label{g2-ghost}
\end{equation}
Torquato et al. \cite{To06d} derived the exact expression for the corresponding structure factor, which is analytic at the origin. The small-$k$ expansion is given by \cite{To06d}
\begin{equation}
    S(k)\sim 0.2901 + 0.01466 k^2+0.004790 k^4, \quad k\rightarrow 0,
\end{equation}
which implies that $v(r)$ has exponential or superexponential decay at large $r$ \cite{To18a}. Thus, we used the functional forms (\ref{hard_core}) and (\ref{exponential_form}) for the basis functions. 

\begin{figure*}[htp]
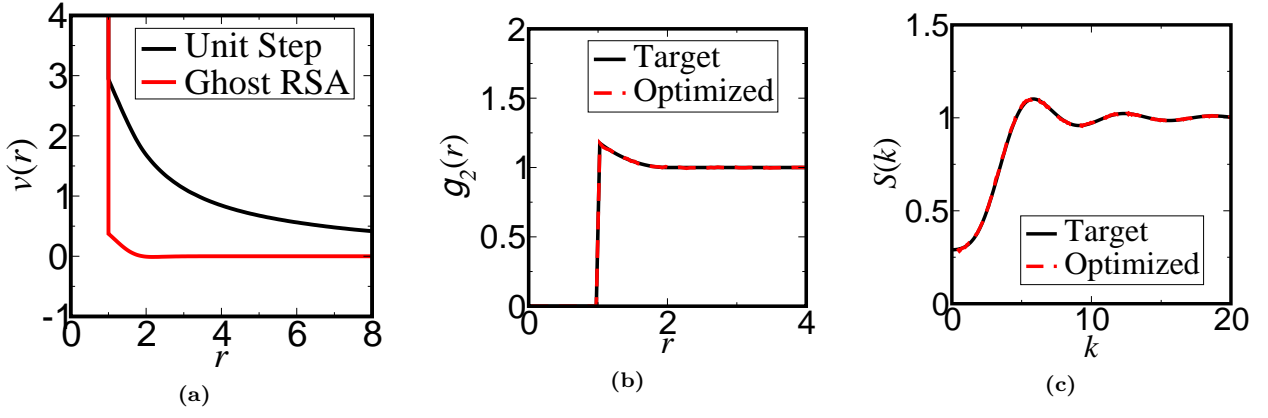

  \centering
  \begin{subfigure}{5cm}
  \includegraphics[width=50mm]{Fig7a.eps}
  \caption{}
  \label{7a}
  \end{subfigure}
  \qquad
  \begin{subfigure}{5cm}
  \includegraphics[width=50mm]{Fig7b.eps}
  \caption{}
  \label{7b}
  \end{subfigure}
  \qquad
  \begin{subfigure}{5cm}
  \includegraphics[width=50mm]{Fig7c.eps}
  \caption{}
  \label{7c}
  \end{subfigure}
  \captionsetup{subrefformat=parens}
  \caption{\subref{7a} The effective potentials for pair statistics of the 3D unit-step-function $g_2$ (\ref{g2-us}) and the 3D ghost RSA process (\ref{g2-ghost}) at  $\phi_c=1/8$. \subref{7b} Target and optimized $g_2(r)$ at $\phi_c$. Here we find that the $L_2$ function (\ref{g2-norm}) is $D_{g_2}=5\times 10^{-4}$. \subref{7c} Target and optimized $S(k)$ at $\phi_c$. Here we find that the $L_2$ function (\ref{S-norm}) is $D_{S}=3\times 10^{-5}$. The $L_2$-norm error (\ref{L2}) is $\mathcal{E}=0.023$.}
  \label{fig:3DRhostRSA}
\end{figure*}

Figure \ref{fig:3DRhostRSA} shows the effective pair potential as well as target and optimized pair statistics for this model. The effective potential is given by
\begin{equation}
        v(r;\phi_c)=
    \begin{cases}
    +\infty, \quad r \leq 1 \\
    \varepsilon_1 \exp\left[-(r/\sigma_1)^2\right] \\ + \varepsilon_2 \exp\left[-(r/\sigma_2)^3\right], \quad r> 1.
    \end{cases}
    \label{v_ghostRSA_3D}
\end{equation}
Table \ref{3DGhostRSA_params} lists the optimized parameters. 
We achieved excellent agreement between target and optimized pair statistics, with the $L_2$ functions [Eqs. (\ref{g2-norm})--(\ref{S-norm})] $D_{g_2}=5\times 10^{-4}$, $D_{S}=3\times 10^{-5}$, and the $L_2$-norm error (\ref{L2}) $\mathcal{E}=0.023$.
Compared to the effective potential for the unit-step-function $g_2$ at the same packing fraction, the potential for the ghost RSA process is much shorter ranged, as the latter system is nonhyperuniform and $S(k)$ is analytic at the origin \cite{To22}. The fact that $v(r;\phi_c)$ for the 3D ghost RSA is not highly repulsive means that the bump in its $g_{2,T}(r)$ does not constitute a considerable constraint to the realizability of the 3D unit-step-function $g_2$ at $\phi_c$.
\begin{center}
\begin{table}
\caption{Optimized parameters of the effective pair potential for the 3D ghost RSA pair statistics.}
\begin{tabular}{ ||c|c|c|c|| } 
 \hline
 $\varepsilon_1$ & $\sigma_1$ & $\varepsilon_2$ & $\sigma_2$\\ 
 \hline
 -0.3502 & 1.284 & 0.8856 & 1.304\\ 
 \hline
\end{tabular}
\label{3DGhostRSA_params}
\end{table}
\end{center}

\section{A known 2D non-realizable target}
\label{impossible}
To test the accuracy and robustness of our methodology, in this section, we study a 2D target $g_2$ whose functional form consists of a unit-step function and a Delta function, separated by a gap \cite{To06b, Zh20}. For specific parameters chosen, the target $g_2$ satisfies all known necessary conditions for realizability (Sec. \ref{nec}), but is not realizable due to geometrical constraints \cite{To06b}. This model has been used to show that the densest packings in high dimensions are disordered \cite{To06b}. 

The functional form of the target $g_2(r)$ is given by
\begin{equation}
    g_{2,T}(r)=\Theta(r-\sigma_T)+\frac{Z}{2\pi\rho}\delta(r-1),
    \label{2DDeltaGap}
\end{equation}
where $\sigma_T\geq 1$ is a distance parameter in units of $\sigma$. When $\sigma_T=1$, $g_{2,T}$ is known as the contact-$\delta$ plus step function \cite{To02c} and is the zero-density limit of the sticky hard sphere (SHS) potential \cite{To02a}
\begin{equation}
    v_{\text{SHS}}(r)=
    \begin{cases}
    +\infty, \quad r<1\\
    -\varepsilon_0, \quad r=1\\
    0, \quad r>1,
    \end{cases}
\end{equation}
which is the limit of the square well (SW) potential \cite{To02a, Sa02}
\begin{equation}
    v_{\text{SW}}(r)=
    \begin{cases}
    +\infty, \quad r<1\\
    -\varepsilon_0, \quad 1\leq r<\sigma_0\\
    0, \quad r\geq \sigma_0,
    \end{cases}
\end{equation}
as $\sigma_0\rightarrow 1$, where $\varepsilon_0$ and $\sigma_0$ are energy and distance parameters in units of $\varepsilon$ and $\sigma$, respectively. The contact-$\delta$ plus step function has been shown to be realizable for a finite range of $r$ for $d=2,3$ \cite{Uc06a}. 

On the other hand, with the parameters $\sigma_T = 1.2946$, $Z=4.0138$ and $\phi=\rho\pi/4=0.74803$, $g_{2,T}(r)$ is not realizable in two dimensions, because of the fact that $Z>4$ implies that there are some particles that are in contact with at least 5 particles. However, any arrangement of the five will result in nonzero $g_2(r)$ within the targeted gap $1<r<1.2946$. 

\begin{figure*}[htp]
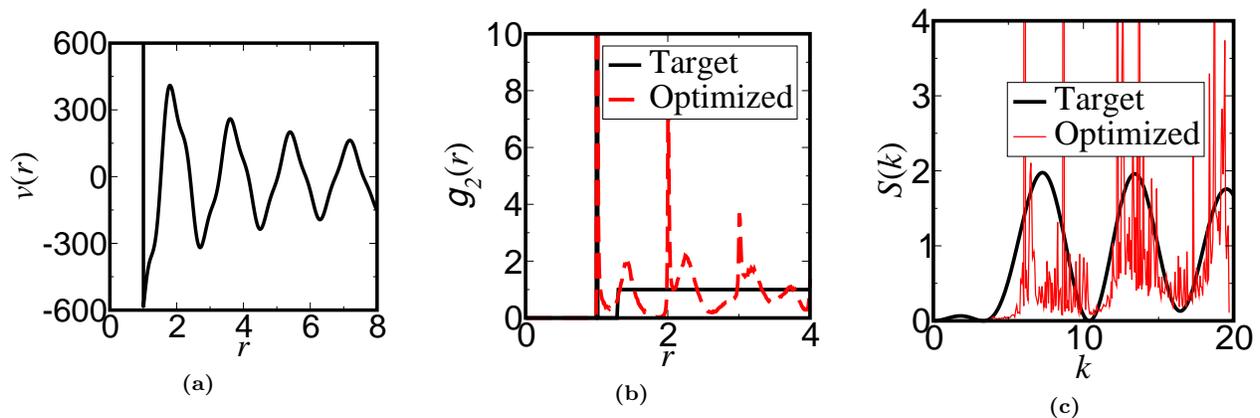

  \centering
  \begin{subfigure}{5cm}
  \includegraphics[width=50mm]{Fig8a.eps}
  \caption{}
  \label{8a}
  \end{subfigure}
  \qquad
  \begin{subfigure}{5cm}
  \includegraphics[width=50mm]{Fig8b.eps}
  \caption{}
  \label{8b}
  \end{subfigure}
  \qquad
  \begin{subfigure}{5cm}
  \includegraphics[width=50mm]{Fig8c.eps}
  \caption{}
  \label{8c}
  \end{subfigure}
  \captionsetup{subrefformat=parens}
  \caption{\subref{8a} The optimized potential for the 2D unrealizable $g_2$ (\ref{2DDeltaGap}), which does not satisfy the convergence criterion. \subref{8b} Target and simulated $g_2(r)$. \subref{8c} Target and simulated $S(k)$.}
  \label{fig:2DDeltaGap}
\end{figure*}

The small-$k$ behavior of the unrealizable target is given by $S_T(k)\sim 0.04689 k^2$, implying that the large-$r$ behavior of the trial $v(r)$ must be of a Coulomb form $v(r)\sim -\log(r)$. The HNC approximation suggests that the small- and intermediate-$r$ behavior of the trial potential is composed of at least 3 oscillatory functions damped by a power-law $r^{-1}$ (\ref{power_form}). The form of the trial potential is thus given by
\begin{equation}
        v(r)=
    \begin{cases}
    +\infty, \quad r \leq 1 \\
    \varepsilon_1\log(r) + \sum_{j=2}^4 \varepsilon_j\cos\left(\frac{r}{\sigma_j} + \theta_j\right)/r, \quad r> 1.
    \end{cases}
    \label{v_impossible}
\end{equation}
As expected, our inverse algorithm did not find an equilibrium state that matches the target pair statistics. The inverse procedure generated an optimized potential [Fig. \ref{fig:2DDeltaGap}(\subref{8a})] that yields $\Psi=59.5$, which is much larger than the convergence criterion $\Psi<5\times 10^{-4}$. The simulated $g_2(r)$ [Fig. \ref{fig:2DDeltaGap}(\subref{8b})] contain extra $\delta$ peaks for $r>1$ and is nonzero in the targeted gap $1<r<1.2946$. The simulated $S(k)$ [Fig. \ref{fig:2DDeltaGap}(\subref{8c})] contain many sharp peaks that are not present in the target. 
We note that the IHNCI procedure for this target yielded diverging trial potentials, i.e., $|v(r)|$ became unbounded for all $r$ values except where $v(r)=0$, which again highlights the insufficiency of such methods for our purpose.

\section{Concluding remarks}
\label{conclusions}
We have numerically investigated the realizability of the unit-step-function $g_2$ for $d=1,2,3$ by determining precise density-dependent effective interactions that yield positive-temperature equilibrium states matching both target $g_2(r)$ for all $r$ and target $S(k)$ for all $k$. 
For $d=1$, we found that the unit-step-function $g_2$ is realizable up to $\phi_m=0.49=0.98\phi_c$. 
For $d=2$ and 3, it is realizable up to the terminal packing fraction $\phi_c=1/2^d$, where the equilibrium states are hyperuniform and $v(r)$ at large $r$ behave as the Coulomb interaction in the respective dimensions. 
This implies that the explicitly known necessary conditions for realizability described in Sec. \ref{nec} are also sufficient up though $\phi_c$ for 2D and 3D unit-step-function $g_2$.
Furthermore, because the unit-step-function $g_2$ is realizable up to $\phi_c$ for $d=2$ and 3, it is also realizable in all higher dimensions up to their respective maximum terminal packing fractions, at which the systems are hyperuniform \cite{To06b, To18a}. 
For $\phi$ near but below $\phi_c$, the large-$r$ behavior of the effective potentials are $\exp[-\kappa(\phi) r]$, $r^{-1/2}\exp[-\kappa(\phi) r]$ and $r^{-1}\exp[-\kappa(\phi) r]$ for $d=1, 2$ and $3$, respectively, where $\kappa(\phi)$ is the inverse screening length, and for $\phi=\phi_c$, the potentials at large $r$ are given by the pure Coulomb forms in the respective dimensions, as predicted in Ref. \citenum{To03a}. 
The effective potentials at the maximum realizable packing fractions for the unit-step-function $g_2$ significantly suppresses both positive and negative correlations that would otherwise be present in the pure equilibrium HS system, as manifested by the smaller order metrics $\tau$ (\ref{tau}) for the corresponding former systems.
The capacity to generate hyperuniform states at positive $T$ via effective potentials is expected to facilitate the self-assembly of tunable hyperuniform soft-matter materials and enables one to probe the thermodynamic and dynamic properties of such states. 
Due to the long-ranged nature of these potentials (\ref{v-long-hu}), one can use $\rho$ and $T$ as tuning parameters to generate equilibrium hyperuniform states or states with stronger hyperuniform forms via their \textit{inherent structures}, i.e. local energy minima \cite{Wa22b}.

While we do not provide a rigorous proof that the 1D unit-step-function $g_2$ is non-realizable at $\phi=\phi_c=1/2$, the accuracy of our methodology and the fact that the hyperuniform targets are realizable in higher dimensions ($d=2$ and $d=3$) suggests that the 1D unit-step-function $g_2$ is highly likely to be non-realizable for $\phi=\phi_c$, even if it is realizable for almost all packing fractions blow $\phi_c$, i.e., for $\phi\le 0.49$.
We recall that it has been shown that the realizability of a particular functional form for hypothetical correlation function corresponding to a disordered system becomes easier as the space dimension increases due to a ``decorrelation'' principle \cite{To06b}, and hence 1D systems are the most likely to be non-realizable in general.

We also found that the pair statistics for the 3D ghost RSA are realizable at the saturation packing fraction $\phi=1/8$. Interestingly, its corresponding effective pair potential is much shorter-ranged than the one we determined for the 3D unit-step-function $g_2$ at $\phi_c$, as the latter system is nonhyperuniform and $S(k)$ is analytic at the origin.
Thus, despite the fact that its saturation density coincides with $\phi_c$ for the unit-step-function $g_2$, the ghost RSA model does not provide a significant constraint on the realizability of the unit-step-function $g_2$.

Importantly, we have demonstrated that our inverse methodology \cite{To22} is robust and generally capable of probing realizability problems, as it successfully generates effective potentials in cases when the target pair statistics is realizable. 
On the other hand, as expected, the trial potential does not converge for the 2D ``contact-Delta + gap + unit-step'' $g_2$ (\ref{2DDeltaGap}) that is known to be non-realizable, despite the fact that it satisfies all known explicit necessary conditions.
Note that the non-realizability of (\ref{2DDeltaGap}) in two dimensions does not imply that it is non-realizable in higher dimensions. A fascinating venue for future research is to apply our methodology to investigate the realizability of (\ref{2DDeltaGap}) for $d\geq 3$. 
If (\ref{2DDeltaGap}) is shown to be realizable in relatively low dimensions, such as $d=3$ or 4, then one has further evidence to the conjecture that the densest packings of identical spheres in high dimensions are disordered (rather than ordered as in low dimensions), which is based on the terminal packing fraction corresponding to the higher-dimensional versions of the ``contact-Delta + gap + unit-step'' $g_2$ \cite{To06b}.

One can also apply our methodology to study the realizability of a wide range of prescribed pair statistics, including $g_2$ with prescribed forms of oscillations \cite{To02c} and hyposurficial states \cite{To03,Zh20}. Comparing the realizable density range for these targets with the terminal densities derived in Ref. \citenum{To02c} might shed light on necessary realizability conditions that are hitherto unknown.

The Zhang-Torquato conjecture implies that the density range on which the step-function $g_2$ is realizable by equilibrium states must coincide with the range on which it is realizable by \textit{any} point process, whether in equilibrium or not.
In the future, an important task will be to investigate the realizability problem by \textit{nonequilibrium} states that match target $g_2(r)$ for all $r$ and target $S(k)$ for all $k$, and to compare the realizable density ranges for equilibrium and nonequilibrium systems. 

Using our inverse procedure, one could also study the thermodynamic and dynamic properties of the effective potentials that yield unit-step and ghost-RSA $g_2$, such as phase diagrams, inherent structures and diffusion properties \cite{St82a, Ha86}.
Unlike ground states and jammed hyperuniform states \cite{To18a, Ma22a}, the particle diffusion rates for the positive-temperature hyperuniform fluids at $\phi_c$ should be positive.
This could occur if the systems' potential only allow neighboring particle pairs to exchange positions. 
Such pair exchanges would allow the system to explore all permutationally equivalent structures, but would prevent other types of structural modifications, thus retaining the system's vanishing isothermal compressibility while permitting a positive self-diffusion rate.
One could test this hypothesis by computing particle trajectories and velocities via molecular dynamics.

Finally, we note that our work provides a challenge to experimentalists to fabricate nanoparticles that result in the effective pair potentials found in this study. 
Achieving the potentials close to (\ref{v_unitstep_2D}) and (\ref{v_unitstep_3D}) for $\phi<\phi_c$ is expected to be experimentally feasible, as the exponentially-damped power-law tails can be readily attained by charged nanoparticles in solution \cite{Ku14}, and the exponential and superexponential short-ranged terms can be achieved by polymer-grafted nanoparticle surfaces \cite{Ha21}. 
Generating interactions close to the ghost RSA potential (\ref{v_ghostRSA_3D}) would also be feasible via polymer-grafted nanoparticles \cite{Ha21}, as (\ref{v_ghostRSA_3D}) consists of only short-ranged superexponential terms. 

If one can find realistic chemical compositions of nanoparticles that realize the unit-step-function $g_2$ near or at the terminal packing fractions, then such systems could be utilized for various practical applications.
For example, nanoparticles interacting via (\ref{v_unitstep_2D}) in the plane near $\phi_c$ suppress clustering, and thus can be useful substrates for reactions that are sensitive to reactant clustering \cite{Tu98}. 
Furthermore, the packings corresponding to the 2D and 3D unit-step-function $g_2$ at $\phi_c$ are hyperuniform states with no short-range order beyond the hard core, unlike typical hyperuniform sphere packings. 
Thus, they can be useful in optical applications \cite{Yu21} and controlled drug delivery \cite{Wa22a}.
While achieving the long-ranged interactions for hyperuniform states is challenging, one could reproduce the potentials over some finite but large range of $r$ to fabricate \textit{effectively hyperuniform} states, i.e., states with very small but nonvanishing positive value of $S(0)$ \cite{He13, At16a, Xu16, Chr17, Mar17}. 
Subsequently, the deviation of such systems from perfect hyperuniformity can be characterized via the various quantitative measures described in Ref. \citenum{To21c}.

\section*{Acknowledgements}
The authors gratefully acknowledge the support of the National Science Foundation under Award No. CBET-2133179. S.T. thanks the Institute for Advanced Study for their hospitality during his sabbatical leave there.


\begin{thebibliography}{76}%
\makeatletter
\providecommand \@ifxundefined [1]{%
 \@ifx{#1\undefined}
}%
\providecommand \@ifnum [1]{%
 \ifnum #1\expandafter \@firstoftwo
 \else \expandafter \@secondoftwo
 \fi
}%
\providecommand \@ifx [1]{%
 \ifx #1\expandafter \@firstoftwo
 \else \expandafter \@secondoftwo
 \fi
}%
\providecommand \natexlab [1]{#1}%
\providecommand \enquote  [1]{``#1''}%
\providecommand \bibnamefont  [1]{#1}%
\providecommand \bibfnamefont [1]{#1}%
\providecommand \citenamefont [1]{#1}%
\providecommand \href@noop [0]{\@secondoftwo}%
\providecommand \href [0]{\begingroup \@sanitize@url \@href}%
\providecommand \@href[1]{\@@startlink{#1}\@@href}%
\providecommand \@@href[1]{\endgroup#1\@@endlink}%
\providecommand \@sanitize@url [0]{\catcode `\\12\catcode `\$12\catcode
  `\&12\catcode `\#12\catcode `\^12\catcode `\_12\catcode `\%12\relax}%
\providecommand \@@startlink[1]{}%
\providecommand \@@endlink[0]{}%
\providecommand \url  [0]{\begingroup\@sanitize@url \@url }%
\providecommand \@url [1]{\endgroup\@href {#1}{\urlprefix }}%
\providecommand \urlprefix  [0]{URL }%
\providecommand \Eprint [0]{\href }%
\providecommand \doibase [0]{https://doi.org/}%
\providecommand \selectlanguage [0]{\@gobble}%
\providecommand \bibinfo  [0]{\@secondoftwo}%
\providecommand \bibfield  [0]{\@secondoftwo}%
\providecommand \translation [1]{[#1]}%
\providecommand \BibitemOpen [0]{}%
\providecommand \bibitemStop [0]{}%
\providecommand \bibitemNoStop [0]{.\EOS\space}%
\providecommand \EOS [0]{\spacefactor3000\relax}%
\providecommand \BibitemShut  [1]{\csname bibitem#1\endcsname}%
\let\auto@bib@innerbib\@empty
\bibitem [{\citenamefont {{Van Workum}}\ and\ \citenamefont
  {Douglas}(2006)}]{Va06}%
  \BibitemOpen
  \bibfield  {author} {\bibinfo {author} {\bibfnamefont {K.}~\bibnamefont {{Van
  Workum}}}\ and\ \bibinfo {author} {\bibfnamefont {J.~F.}\ \bibnamefont
  {Douglas}},\ }\bibfield  {title} {\enquote {\bibinfo {title} {{Symmetry,
  equivalence, and molecular self-assembly}},}\ }\href
  {https://doi.org/10.1103/PhysRevE.73.031502} {\bibfield  {journal} {\bibinfo
  {journal} {Phys. Rev. E}\ }\textbf {\bibinfo {volume} {73}},\ \bibinfo
  {pages} {031502} (\bibinfo {year} {2006})}\BibitemShut {NoStop}%
\bibitem [{\citenamefont {Rechtsman}, \citenamefont {Stillinger},\ and\
  \citenamefont {Torquato}(2006)}]{Re06a}%
  \BibitemOpen
  \bibfield  {author} {\bibinfo {author} {\bibfnamefont {M.~C.}\ \bibnamefont
  {Rechtsman}}, \bibinfo {author} {\bibfnamefont {F.~H.}\ \bibnamefont
  {Stillinger}},\ and\ \bibinfo {author} {\bibfnamefont {S.}~\bibnamefont
  {Torquato}},\ }\bibfield  {title} {\enquote {\bibinfo {title} {Designed
  isotropic potentials via inverse methods for self-assembly},}\ }\href
  {https://doi.org/10.1103/PhysRevE.73.011406} {\bibfield  {journal} {\bibinfo
  {journal} {Phys. Rev. E}\ }\textbf {\bibinfo {volume} {73}},\ \bibinfo
  {pages} {011406:1--12} (\bibinfo {year} {2006})},\ \bibinfo {note}
  {\href{https://journals.aps.org/pre/abstract/10.1103/PhysRevE.75.019902}{{E}rratum,
  {\bf 75}, 019902(E) {(}2006{)}}.}\BibitemShut {Stop}%
\bibitem [{\citenamefont {Torquato}(2009)}]{To09a}%
  \BibitemOpen
  \bibfield  {author} {\bibinfo {author} {\bibfnamefont {S.}~\bibnamefont
  {Torquato}},\ }\bibfield  {title} {\enquote {\bibinfo {title} {Inverse
  optimization techniques for targeted self-assembly},}\ }\href
  {https://doi.org/10.1039/B814211B} {\bibfield  {journal} {\bibinfo  {journal}
  {Soft Matter}\ }\textbf {\bibinfo {volume} {5}},\ \bibinfo {pages}
  {1157--1173} (\bibinfo {year} {2009})}\BibitemShut {NoStop}%
\bibitem [{\citenamefont {Cohn}\ and\ \citenamefont {Kumar}(2009)}]{Co09}%
  \BibitemOpen
  \bibfield  {author} {\bibinfo {author} {\bibfnamefont {H.}~\bibnamefont
  {Cohn}}\ and\ \bibinfo {author} {\bibfnamefont {A.}~\bibnamefont {Kumar}},\
  }\bibfield  {title} {\enquote {\bibinfo {title} {Optimality and uniqueness of
  the leech lattice among lattices},}\ }\href
  {https://doi.org/10.4007/annals.2009.170.1003} {\bibfield  {journal}
  {\bibinfo  {journal} {Ann. Math.}\ }\textbf {\bibinfo {volume} {170}},\
  \bibinfo {pages} {1003--1050} (\bibinfo {year} {2009})}\BibitemShut {NoStop}%
\bibitem [{\citenamefont {Torquato}(2002)}]{To02a}%
  \BibitemOpen
  \bibfield  {author} {\bibinfo {author} {\bibfnamefont {S.}~\bibnamefont
  {Torquato}},\ }\href@noop {} {\emph {\bibinfo {title} {Random Heterogeneous
  Materials: Microstructure and Macroscopic Properties}}}\ (\bibinfo
  {publisher} {Springer-Verlag},\ \bibinfo {address} {New York},\ \bibinfo
  {year} {2002})\BibitemShut {NoStop}%
\bibitem [{\citenamefont {Hansen}\ and\ \citenamefont {McDonald}(1986)}]{Ha86}%
  \BibitemOpen
  \bibfield  {author} {\bibinfo {author} {\bibfnamefont {J.~P.}\ \bibnamefont
  {Hansen}}\ and\ \bibinfo {author} {\bibfnamefont {I.~R.}\ \bibnamefont
  {McDonald}},\ }\href@noop {} {\emph {\bibinfo {title} {Theory of Simple
  Liquids}}}\ (\bibinfo  {publisher} {Academic Press},\ \bibinfo {address} {New
  York},\ \bibinfo {year} {1986})\BibitemShut {NoStop}%
\bibitem [{\citenamefont {Yamada}(1961)}]{Ya61}%
  \BibitemOpen
  \bibfield  {author} {\bibinfo {author} {\bibfnamefont {M.}~\bibnamefont
  {Yamada}},\ }\bibfield  {title} {\enquote {\bibinfo {title} {Geometrical
  study of the pair distribution function in the many-body problem},}\ }\href
  {https://doi.org/10.1143/PTP.25.579} {\bibfield  {journal} {\bibinfo
  {journal} {Prog. Theor. Phys.}\ }\textbf {\bibinfo {volume} {25}},\ \bibinfo
  {pages} {579--594} (\bibinfo {year} {1961})}\BibitemShut {NoStop}%
\bibitem [{\citenamefont {Crawford}, \citenamefont {Torquato},\ and\
  \citenamefont {Stillinger}(2003)}]{Cr03}%
  \BibitemOpen
  \bibfield  {author} {\bibinfo {author} {\bibfnamefont {J.~R.}\ \bibnamefont
  {Crawford}}, \bibinfo {author} {\bibfnamefont {S.}~\bibnamefont {Torquato}},\
  and\ \bibinfo {author} {\bibfnamefont {F.~H.}\ \bibnamefont {Stillinger}},\
  }\bibfield  {title} {\enquote {\bibinfo {title} {Aspects of correlation
  function realizability},}\ }\href {https://doi.org/10.1063/1.1606678}
  {\bibfield  {journal} {\bibinfo  {journal} {J. Chem. Phys.}\ }\textbf
  {\bibinfo {volume} {119}},\ \bibinfo {pages} {7065--7074} (\bibinfo {year}
  {2003})}\BibitemShut {NoStop}%
\bibitem [{\citenamefont {Costin}\ and\ \citenamefont
  {Lebowitz}(2004)}]{Cos04}%
  \BibitemOpen
  \bibfield  {author} {\bibinfo {author} {\bibfnamefont {O.}~\bibnamefont
  {Costin}}\ and\ \bibinfo {author} {\bibfnamefont {J.}~\bibnamefont
  {Lebowitz}},\ }\bibfield  {title} {\enquote {\bibinfo {title} {On the
  construction of particle distributions with specified single and pair
  densities},}\ }\href {https://doi.org/10.1021/jp047793m} {\bibfield
  {journal} {\bibinfo  {journal} {J. Phys. Chem. B.}\ }\textbf {\bibinfo
  {volume} {108}},\ \bibinfo {pages} {19614--19618} (\bibinfo {year}
  {2004})}\BibitemShut {NoStop}%
\bibitem [{\citenamefont {Uche}, \citenamefont {Stillinger},\ and\
  \citenamefont {Torquato}(2006)}]{Uc06a}%
  \BibitemOpen
  \bibfield  {author} {\bibinfo {author} {\bibfnamefont {O.~U.}\ \bibnamefont
  {Uche}}, \bibinfo {author} {\bibfnamefont {F.~H.}\ \bibnamefont
  {Stillinger}},\ and\ \bibinfo {author} {\bibfnamefont {S.}~\bibnamefont
  {Torquato}},\ }\bibfield  {title} {\enquote {\bibinfo {title} {On the
  realizability of pair correlation functions},}\ }\href
  {https://doi.org/10.1016/j.physa.2005.03.058} {\bibfield  {journal} {\bibinfo
   {journal} {Physica A}\ }\textbf {\bibinfo {volume} {360}},\ \bibinfo {pages}
  {21--36} (\bibinfo {year} {2006})}\BibitemShut {NoStop}%
\bibitem [{\citenamefont {Torquato}\ and\ \citenamefont
  {Stillinger}(2006{\natexlab{a}})}]{To06b}%
  \BibitemOpen
  \bibfield  {author} {\bibinfo {author} {\bibfnamefont {S.}~\bibnamefont
  {Torquato}}\ and\ \bibinfo {author} {\bibfnamefont {F.~H.}\ \bibnamefont
  {Stillinger}},\ }\bibfield  {title} {\enquote {\bibinfo {title} {New
  conjectural lower bounds on the optimal density of sphere packings},}\ }\href
  {https://doi.org/10.1080/10586458.2006.10128964} {\bibfield  {journal}
  {\bibinfo  {journal} {Experimental Math.}\ }\textbf {\bibinfo {volume}
  {15}},\ \bibinfo {pages} {307--331} (\bibinfo {year}
  {2006}{\natexlab{a}})}\BibitemShut {NoStop}%
\bibitem [{\citenamefont {Kuna}, \citenamefont {Lebowitz},\ and\ \citenamefont
  {Speer}(2007)}]{Ku07}%
  \BibitemOpen
  \bibfield  {author} {\bibinfo {author} {\bibfnamefont {T.}~\bibnamefont
  {Kuna}}, \bibinfo {author} {\bibfnamefont {J.~L.}\ \bibnamefont {Lebowitz}},\
  and\ \bibinfo {author} {\bibfnamefont {E.~R.}\ \bibnamefont {Speer}},\
  }\bibfield  {title} {\enquote {\bibinfo {title} {Realizability of point
  processes},}\ }\href {https://doi.org/10.1007/s10955-007-9393-y} {\bibfield
  {journal} {\bibinfo  {journal} {J. Stat. Phys.}\ }\textbf {\bibinfo {volume}
  {129}},\ \bibinfo {pages} {417--439} (\bibinfo {year} {2007})}\BibitemShut
  {NoStop}%
\bibitem [{\citenamefont {Zhang}\ and\ \citenamefont {Torquato}(2020)}]{Zh20}%
  \BibitemOpen
  \bibfield  {author} {\bibinfo {author} {\bibfnamefont {G.}~\bibnamefont
  {Zhang}}\ and\ \bibinfo {author} {\bibfnamefont {S.}~\bibnamefont
  {Torquato}},\ }\bibfield  {title} {\enquote {\bibinfo {title} {Realizable
  hyperuniform and nonhyperuniform particle configurations with targeted
  spectral functions via effective pair interactions},}\ }\href
  {https://doi.org/10.1103/PhysRevE.101.032124} {\bibfield  {journal} {\bibinfo
   {journal} {Phys. Rev. E}\ }\textbf {\bibinfo {volume} {101}},\ \bibinfo
  {pages} {032124} (\bibinfo {year} {2020})}\BibitemShut {NoStop}%
\bibitem [{\citenamefont {Torquato}\ and\ \citenamefont
  {Stillinger}(2002)}]{To02c}%
  \BibitemOpen
  \bibfield  {author} {\bibinfo {author} {\bibfnamefont {S.}~\bibnamefont
  {Torquato}}\ and\ \bibinfo {author} {\bibfnamefont {F.~H.}\ \bibnamefont
  {Stillinger}},\ }\bibfield  {title} {\enquote {\bibinfo {title} {Controlling
  the short-range order and packing densities of many-particle systems},}\
  }\href {https://doi.org/10.1021/jp0208687} {\bibfield  {journal} {\bibinfo
  {journal} {J. Phys. Chem. B}\ }\textbf {\bibinfo {volume} {106}},\ \bibinfo
  {pages} {8354--8359} (\bibinfo {year} {2002})},\ \bibinfo {note}
  {\href{https://pubs.acs.org/doi/10.1021/jp022019p}{{E}rratum {\bf 106}, 11406
  (2002).}}\BibitemShut {Stop}%
\bibitem [{\citenamefont {Torquato}\ and\ \citenamefont
  {Stillinger}(2003)}]{To03a}%
  \BibitemOpen
  \bibfield  {author} {\bibinfo {author} {\bibfnamefont {S.}~\bibnamefont
  {Torquato}}\ and\ \bibinfo {author} {\bibfnamefont {F.~H.}\ \bibnamefont
  {Stillinger}},\ }\bibfield  {title} {\enquote {\bibinfo {title} {Local
  density fluctuations, hyperuniform systems, and order metrics},}\ }\href
  {https://doi.org/10.1103/PhysRevE.68.041113} {\bibfield  {journal} {\bibinfo
  {journal} {Phys. Rev. E}\ }\textbf {\bibinfo {volume} {68}},\ \bibinfo
  {pages} {041113} (\bibinfo {year} {2003})}\BibitemShut {NoStop}%
\bibitem [{\citenamefont {Stillinger}\ and\ \citenamefont
  {Torquato}(2005)}]{St05}%
  \BibitemOpen
  \bibfield  {author} {\bibinfo {author} {\bibfnamefont {F.~H.}\ \bibnamefont
  {Stillinger}}\ and\ \bibinfo {author} {\bibfnamefont {S.}~\bibnamefont
  {Torquato}},\ }\bibfield  {title} {\enquote {\bibinfo {title} {Realizability
  issues for iso-$g^{(2)}$ processes},}\ }\href
  {https://doi.org/10.1080/00268970500151528} {\bibfield  {journal} {\bibinfo
  {journal} {Mol. Phys.}\ }\textbf {\bibinfo {volume} {103}},\ \bibinfo {pages}
  {2943--2949} (\bibinfo {year} {2005})}\BibitemShut {NoStop}%
\bibitem [{\citenamefont {Torquato}(2018{\natexlab{a}})}]{To18a}%
  \BibitemOpen
  \bibfield  {author} {\bibinfo {author} {\bibfnamefont {S.}~\bibnamefont
  {Torquato}},\ }\bibfield  {title} {\enquote {\bibinfo {title} {Hyperuniform
  states of matter},}\ }\href {https://doi.org/10.1016/j.physrep.2018.03.001}
  {\bibfield  {journal} {\bibinfo  {journal} {Physics Reports}\ }\textbf
  {\bibinfo {volume} {745}},\ \bibinfo {pages} {1--95} (\bibinfo {year}
  {2018}{\natexlab{a}})}\BibitemShut {NoStop}%
\bibitem [{\citenamefont {Stillinger}\ \emph {et~al.}(2001)\citenamefont
  {Stillinger}, \citenamefont {Torquato}, \citenamefont {Eroles},\ and\
  \citenamefont {Truskett}}]{St01a}%
  \BibitemOpen
  \bibfield  {author} {\bibinfo {author} {\bibfnamefont {F.~H.}\ \bibnamefont
  {Stillinger}}, \bibinfo {author} {\bibfnamefont {S.}~\bibnamefont
  {Torquato}}, \bibinfo {author} {\bibfnamefont {J.~M.}\ \bibnamefont
  {Eroles}},\ and\ \bibinfo {author} {\bibfnamefont {T.~M.}\ \bibnamefont
  {Truskett}},\ }\bibfield  {title} {\enquote {\bibinfo {title} {Iso-$g^{(2)}$
  processes in equilibrium statistical mechanics},}\ }\href
  {https://doi.org/10.1021/jp010006q} {\bibfield  {journal} {\bibinfo
  {journal} {J. Phys. Chem. B}\ }\textbf {\bibinfo {volume} {105}},\ \bibinfo
  {pages} {6592--6597} (\bibinfo {year} {2001})}\BibitemShut {NoStop}%
\bibitem [{\citenamefont {McGreevy}(2001)}]{Mc01}%
  \BibitemOpen
  \bibfield  {author} {\bibinfo {author} {\bibfnamefont {R.~L.}\ \bibnamefont
  {McGreevy}},\ }\bibfield  {title} {\enquote {\bibinfo {title} {Reverse monte
  carlo modelling},}\ }\href {https://doi.org/10.1088/0953-8984/13/46/201}
  {\bibfield  {journal} {\bibinfo  {journal} {Journal of Physics: Condensed
  Matter}\ }\textbf {\bibinfo {volume} {13}},\ \bibinfo {pages} {R877--R913}
  (\bibinfo {year} {2001})}\BibitemShut {NoStop}%
\bibitem [{\citenamefont {Levesque}, \citenamefont {Weis},\ and\ \citenamefont
  {Reatto}(1985)}]{Le85}%
  \BibitemOpen
  \bibfield  {author} {\bibinfo {author} {\bibfnamefont {D.}~\bibnamefont
  {Levesque}}, \bibinfo {author} {\bibfnamefont {J.~J.}\ \bibnamefont {Weis}},\
  and\ \bibinfo {author} {\bibfnamefont {L.}~\bibnamefont {Reatto}},\
  }\bibfield  {title} {\enquote {\bibinfo {title} {Pair interaction from
  structural data for dense classical liquids},}\ }\href
  {https://doi.org/10.1103/PhysRevLett.54.451} {\bibfield  {journal} {\bibinfo
  {journal} {Phys. Rev. Lett.}\ }\textbf {\bibinfo {volume} {54}},\ \bibinfo
  {pages} {451--454} (\bibinfo {year} {1985})}\BibitemShut {NoStop}%
\bibitem [{\citenamefont {Lyubartsev}\ and\ \citenamefont
  {Laaksonen}(1995)}]{Ly95}%
  \BibitemOpen
  \bibfield  {author} {\bibinfo {author} {\bibfnamefont {A.~P.}\ \bibnamefont
  {Lyubartsev}}\ and\ \bibinfo {author} {\bibfnamefont {A.}~\bibnamefont
  {Laaksonen}},\ }\bibfield  {title} {\enquote {\bibinfo {title} {Calculation
  of effective interaction potentials from radial distribution functions: A
  reverse monte carlo approach},}\ }\href
  {https://doi.org/10.1103/PhysRevE.52.3730} {\bibfield  {journal} {\bibinfo
  {journal} {Phys. Rev. E}\ }\textbf {\bibinfo {volume} {52}},\ \bibinfo
  {pages} {3730--3737} (\bibinfo {year} {1995})}\BibitemShut {NoStop}%
\bibitem [{\citenamefont {Soper}(1996)}]{So96}%
  \BibitemOpen
  \bibfield  {author} {\bibinfo {author} {\bibfnamefont {A.~K.}\ \bibnamefont
  {Soper}},\ }\bibfield  {title} {\enquote {\bibinfo {title} {{Empirical
  potential Monte Carlo simulation of fluid structure}},}\ }\href
  {https://doi.org/10.1016/0301-0104(95)00357-6} {\bibfield  {journal}
  {\bibinfo  {journal} {Chem. Phys.}\ }\textbf {\bibinfo {volume} {202}},\
  \bibinfo {pages} {295--306} (\bibinfo {year} {1996})}\BibitemShut {NoStop}%
\bibitem [{\citenamefont {Jain}, \citenamefont {Errington},\ and\ \citenamefont
  {Truskett}(2013)}]{Jain13}%
  \BibitemOpen
  \bibfield  {author} {\bibinfo {author} {\bibfnamefont {A.}~\bibnamefont
  {Jain}}, \bibinfo {author} {\bibfnamefont {J.~R.}\ \bibnamefont
  {Errington}},\ and\ \bibinfo {author} {\bibfnamefont {T.~M.}\ \bibnamefont
  {Truskett}},\ }\bibfield  {title} {\enquote {\bibinfo {title} {Inverse design
  of simple pairwise interactions with low-coordinated 3d lattice ground
  states},}\ }\href {https://doi.org/10.1063/1.5063802} {\bibfield  {journal}
  {\bibinfo  {journal} {Soft Matter}\ }\textbf {\bibinfo {volume} {9}},\
  \bibinfo {pages} {3866--3870} (\bibinfo {year} {2013})}\BibitemShut {NoStop}%
\bibitem [{\citenamefont {Heinen}(2018)}]{He18}%
  \BibitemOpen
  \bibfield  {author} {\bibinfo {author} {\bibfnamefont {M.}~\bibnamefont
  {Heinen}},\ }\bibfield  {title} {\enquote {\bibinfo {title} {{Calculating
  particle pair potentials from fluid-state pair correlations: Iterative
  {O}rnstein–{Z}ernike inversion}},}\ }\href
  {https://doi.org/10.1002/jcc.25225} {\bibfield  {journal} {\bibinfo
  {journal} {J. Comput. Chem.}\ }\textbf {\bibinfo {volume} {39}},\ \bibinfo
  {pages} {1531--1543} (\bibinfo {year} {2018})}\BibitemShut {NoStop}%
\bibitem [{\citenamefont {Torquato}\ and\ \citenamefont {Wang}(2022)}]{To22}%
  \BibitemOpen
  \bibfield  {author} {\bibinfo {author} {\bibfnamefont {S.}~\bibnamefont
  {Torquato}}\ and\ \bibinfo {author} {\bibfnamefont {H.}~\bibnamefont
  {Wang}},\ }\bibfield  {title} {\enquote {\bibinfo {title} {Precise
  determination of pair interactions from pair statistics of many-body systems
  in and out of equilibrium},}\ }\href
  {https://doi.org/10.1103/PhysRevE.106.044122} {\bibfield  {journal} {\bibinfo
   {journal} {Phys. Rev. E}\ }\textbf {\bibinfo {volume} {106}},\ \bibinfo
  {pages} {044122} (\bibinfo {year} {2022})}\BibitemShut {NoStop}%
\bibitem [{\citenamefont {Wang}\ and\ \citenamefont
  {Torquato}(2022{\natexlab{a}})}]{Wa22b}%
  \BibitemOpen
  \bibfield  {author} {\bibinfo {author} {\bibfnamefont {H.}~\bibnamefont
  {Wang}}\ and\ \bibinfo {author} {\bibfnamefont {S.}~\bibnamefont
  {Torquato}},\ }\bibfield  {title} {\enquote {\bibinfo {title}
  {\href{https://arxiv.org/abs/2210.06446}{Equilibrium States Corresponding to
  Targeted Hyperuniform Nonequilibrium Pair Statistics}},}\ }\href@noop {}
  {\bibfield  {journal} {\bibinfo  {journal} {In review}\ } (\bibinfo {year}
  {2022}{\natexlab{a}})}\BibitemShut {NoStop}%
\bibitem [{\citenamefont {Feder}(1980)}]{Fe80}%
  \BibitemOpen
  \bibfield  {author} {\bibinfo {author} {\bibfnamefont {J.}~\bibnamefont
  {Feder}},\ }\bibfield  {title} {\enquote {\bibinfo {title} {Random sequential
  adsorption},}\ }\href {https://doi.org/10.1016/0022-5193(80)90358-6}
  {\bibfield  {journal} {\bibinfo  {journal} {J. Theor. Biol.}\ }\textbf
  {\bibinfo {volume} {87}},\ \bibinfo {pages} {237--254} (\bibinfo {year}
  {1980})}\BibitemShut {NoStop}%
\bibitem [{\citenamefont {Cooper}(1988)}]{Co88}%
  \BibitemOpen
  \bibfield  {author} {\bibinfo {author} {\bibfnamefont {D.~W.}\ \bibnamefont
  {Cooper}},\ }\bibfield  {title} {\enquote {\bibinfo {title}
  {Random-sequential-packing simulations in three dimensions for spheres},}\
  }\href {https://doi.org/10.1103/PhysRevA.38.522} {\bibfield  {journal}
  {\bibinfo  {journal} {Phys. Rev. A}\ }\textbf {\bibinfo {volume} {38}},\
  \bibinfo {pages} {522--524} (\bibinfo {year} {1988})}\BibitemShut {NoStop}%
\bibitem [{\citenamefont {Torquato}\ and\ \citenamefont
  {Stillinger}(2006{\natexlab{b}})}]{To06a}%
  \BibitemOpen
  \bibfield  {author} {\bibinfo {author} {\bibfnamefont {S.}~\bibnamefont
  {Torquato}}\ and\ \bibinfo {author} {\bibfnamefont {F.~H.}\ \bibnamefont
  {Stillinger}},\ }\bibfield  {title} {\enquote {\bibinfo {title} {Exactly
  solvable disordered sphere-packing model in arbitrary-dimensional {E}uclidean
  spaces},}\ }\href {https://doi.org/10.1103/PhysRevE.73.031106} {\bibfield
  {journal} {\bibinfo  {journal} {Phys. Rev. E}\ }\textbf {\bibinfo {volume}
  {73}},\ \bibinfo {pages} {031106} (\bibinfo {year}
  {2006}{\natexlab{b}})}\BibitemShut {NoStop}%
\bibitem [{\citenamefont {Zhang}\ and\ \citenamefont {Torquato}(2013)}]{Zh13b}%
  \BibitemOpen
  \bibfield  {author} {\bibinfo {author} {\bibfnamefont {G.}~\bibnamefont
  {Zhang}}\ and\ \bibinfo {author} {\bibfnamefont {S.}~\bibnamefont
  {Torquato}},\ }\bibfield  {title} {\enquote {\bibinfo {title} {Precise
  algorithm to generate random sequential addition of hard hyperspheres at
  saturation},}\ }\href {https://doi.org/10.1103/PhysRevE.88.053312} {\bibfield
   {journal} {\bibinfo  {journal} {Phys. Rev. E}\ }\textbf {\bibinfo {volume}
  {88}},\ \bibinfo {pages} {053312} (\bibinfo {year} {2013})}\BibitemShut
  {NoStop}%
\bibitem [{\citenamefont {Zhang}, \citenamefont {Stillinger},\ and\
  \citenamefont {Torquato}(2016)}]{Zh16a}%
  \BibitemOpen
  \bibfield  {author} {\bibinfo {author} {\bibfnamefont {G.}~\bibnamefont
  {Zhang}}, \bibinfo {author} {\bibfnamefont {F.~H.}\ \bibnamefont
  {Stillinger}},\ and\ \bibinfo {author} {\bibfnamefont {S.}~\bibnamefont
  {Torquato}},\ }\bibfield  {title} {\enquote {\bibinfo {title} {The perfect
  glass paradigm: Disordered hyperuniform glasses down to absolute zero},}\
  }\href {https://doi.org/10.1038/srep36963} {\bibfield  {journal} {\bibinfo
  {journal} {Sci. Rep.}\ }\textbf {\bibinfo {volume} {6}},\ \bibinfo {pages}
  {36963} (\bibinfo {year} {2016})}\BibitemShut {NoStop}%
\bibitem [{\citenamefont {Klatt}, \citenamefont {Kim},\ and\ \citenamefont
  {Torquato}(2020)}]{Kl20}%
  \BibitemOpen
  \bibfield  {author} {\bibinfo {author} {\bibfnamefont {M.~A.}\ \bibnamefont
  {Klatt}}, \bibinfo {author} {\bibfnamefont {J.}~\bibnamefont {Kim}},\ and\
  \bibinfo {author} {\bibfnamefont {S.}~\bibnamefont {Torquato}},\ }\bibfield
  {title} {\enquote {\bibinfo {title} {Cloaking the underlying long-range order
  of randomly perturbed lattices},}\ }\href
  {https://doi.org/10.1103/PhysRevE.101.032118} {\bibfield  {journal} {\bibinfo
   {journal} {Phys. Rev. E}\ }\textbf {\bibinfo {volume} {101}},\ \bibinfo
  {pages} {032118} (\bibinfo {year} {2020})}\BibitemShut {NoStop}%
\bibitem [{\citenamefont {Cort{\'{e}}}\ \emph {et~al.}(2008)\citenamefont
  {Cort{\'{e}}}, \citenamefont {Chaikin}, \citenamefont {Gollub},\ and\
  \citenamefont {Pine}}]{Co08}%
  \BibitemOpen
  \bibfield  {author} {\bibinfo {author} {\bibfnamefont {L.}~\bibnamefont
  {Cort{\'{e}}}}, \bibinfo {author} {\bibfnamefont {P.~M.}\ \bibnamefont
  {Chaikin}}, \bibinfo {author} {\bibfnamefont {J.~P.}\ \bibnamefont
  {Gollub}},\ and\ \bibinfo {author} {\bibfnamefont {D.~J.}\ \bibnamefont
  {Pine}},\ }\bibfield  {title} {\enquote {\bibinfo {title} {{Random
  organization in periodically driven systems}},}\ }\href
  {https://doi.org/10.1038/nphys891} {\bibfield  {journal} {\bibinfo  {journal}
  {Nat. Phys.}\ }\textbf {\bibinfo {volume} {4}},\ \bibinfo {pages} {420--424}
  (\bibinfo {year} {2008})}\BibitemShut {NoStop}%
\bibitem [{\citenamefont {{Hexner}}\ and\ \citenamefont
  {{Levine}}(2015)}]{He15}%
  \BibitemOpen
  \bibfield  {author} {\bibinfo {author} {\bibfnamefont {D.}~\bibnamefont
  {{Hexner}}}\ and\ \bibinfo {author} {\bibfnamefont {D.}~\bibnamefont
  {{Levine}}},\ }\bibfield  {title} {\enquote {\bibinfo {title}
  {{Hyperuniformity of critical absorbing states}},}\ }\href
  {https://doi.org/10.1103/PhysRevLett.114.110602} {\bibfield  {journal}
  {\bibinfo  {journal} {Phys. Rev. Lett.}\ }\textbf {\bibinfo {volume} {114}},\
  \bibinfo {pages} {110602} (\bibinfo {year} {2015})}\BibitemShut {NoStop}%
\bibitem [{\citenamefont {Sakai}, \citenamefont {Stillinger},\ and\
  \citenamefont {Torquato}(2002)}]{Sa02}%
  \BibitemOpen
  \bibfield  {author} {\bibinfo {author} {\bibfnamefont {H.}~\bibnamefont
  {Sakai}}, \bibinfo {author} {\bibfnamefont {F.~H.}\ \bibnamefont
  {Stillinger}},\ and\ \bibinfo {author} {\bibfnamefont {S.}~\bibnamefont
  {Torquato}},\ }\bibfield  {title} {\enquote {\bibinfo {title} {Equi-$g(r)$
  sequences of systems derived from the square-well potential},}\ }\href
  {https://doi.org/10.1063/1.1480864} {\bibfield  {journal} {\bibinfo
  {journal} {J. Chem. Phys.}\ }\textbf {\bibinfo {volume} {117}},\ \bibinfo
  {pages} {297--307} (\bibinfo {year} {2002})}\BibitemShut {NoStop}%
\bibitem [{\citenamefont {Stillinger}, \citenamefont {Sakai},\ and\
  \citenamefont {Torquato}(2002)}]{st02}%
  \BibitemOpen
  \bibfield  {author} {\bibinfo {author} {\bibfnamefont {F.~H.}\ \bibnamefont
  {Stillinger}}, \bibinfo {author} {\bibfnamefont {H.}~\bibnamefont {Sakai}},\
  and\ \bibinfo {author} {\bibfnamefont {S.}~\bibnamefont {Torquato}},\
  }\bibfield  {title} {\enquote {\bibinfo {title} {Statistical mechanical
  models with effective potentials: Definitions, applications, and
  thermodynamic consequences},}\ }\href {https://doi.org/10.1063/1.1480863}
  {\bibfield  {journal} {\bibinfo  {journal} {J. Chem. Phys.}\ }\textbf
  {\bibinfo {volume} {117}},\ \bibinfo {pages} {288--296} (\bibinfo {year}
  {2002})}\BibitemShut {NoStop}%
\bibitem [{\citenamefont {Levine}\ and\ \citenamefont
  {Steinhardt}(1984)}]{Le84}%
  \BibitemOpen
  \bibfield  {author} {\bibinfo {author} {\bibfnamefont {D.}~\bibnamefont
  {Levine}}\ and\ \bibinfo {author} {\bibfnamefont {P.~J.}\ \bibnamefont
  {Steinhardt}},\ }\bibfield  {title} {\enquote {\bibinfo {title}
  {Quasicrystals: {A} new class of ordered structures},}\ }\href@noop {}
  {\bibfield  {journal} {\bibinfo  {journal} {Phys. Rev. Lett.}\ }\textbf
  {\bibinfo {volume} {53}},\ \bibinfo {pages} {2477--2480} (\bibinfo {year}
  {1984})}\BibitemShut {NoStop}%
\bibitem [{\citenamefont {Torquato}\ and\ \citenamefont
  {Stillinger}(2008)}]{To08a}%
  \BibitemOpen
  \bibfield  {author} {\bibinfo {author} {\bibfnamefont {S.}~\bibnamefont
  {Torquato}}\ and\ \bibinfo {author} {\bibfnamefont {F.~H.}\ \bibnamefont
  {Stillinger}},\ }\bibfield  {title} {\enquote {\bibinfo {title} {New duality
  relations for classical ground states,},}\ }\href@noop {} {\bibfield
  {journal} {\bibinfo  {journal} {Phys. Rev. Lett.}\ }\textbf {\bibinfo
  {volume} {100}},\ \bibinfo {pages} {020602} (\bibinfo {year}
  {2008})}\BibitemShut {NoStop}%
\bibitem [{\citenamefont {Henderson}(1974)}]{He74}%
  \BibitemOpen
  \bibfield  {author} {\bibinfo {author} {\bibfnamefont {R.~L.}\ \bibnamefont
  {Henderson}},\ }\bibfield  {title} {\enquote {\bibinfo {title} {{A uniqueness
  theorem for fluid pair correlation functions}},}\ }\href@noop {} {\bibfield
  {journal} {\bibinfo  {journal} {Phys. Lett. A}\ }\textbf {\bibinfo {volume}
  {49}},\ \bibinfo {pages} {197--198} (\bibinfo {year} {1974})}\BibitemShut
  {NoStop}%
\bibitem [{\citenamefont {Re{\'n}yi}(1963)}]{Re63}%
  \BibitemOpen
  \bibfield  {author} {\bibinfo {author} {\bibfnamefont {A.}~\bibnamefont
  {Re{\'n}yi}},\ }\bibfield  {title} {\enquote {\bibinfo {title} {On a
  one-dimensional problem concerning random space filling},}\ }\href
  {https://doi.org/doi.org/10.2307/3212263} {\bibfield  {journal} {\bibinfo
  {journal} {Sel. Trans. Math. Stat. Prob.}\ }\textbf {\bibinfo {volume} {4}},\
  \bibinfo {pages} {203--218} (\bibinfo {year} {1963})}\BibitemShut {NoStop}%
\bibitem [{\citenamefont {Widom}(1966)}]{wi66}%
  \BibitemOpen
  \bibfield  {author} {\bibinfo {author} {\bibfnamefont {B.}~\bibnamefont
  {Widom}},\ }\bibfield  {title} {\enquote {\bibinfo {title} {Random sequential
  addition of hard spheres to a volume},}\ }\href
  {https://doi.org/10.1063/1.1726548} {\bibfield  {journal} {\bibinfo
  {journal} {J. Chem. Phys.}\ }\textbf {\bibinfo {volume} {44}},\ \bibinfo
  {pages} {3888--3894} (\bibinfo {year} {1966})}\BibitemShut {NoStop}%
\bibitem [{\citenamefont {Zachary}\ and\ \citenamefont
  {Torquato}(2009)}]{Za09}%
  \BibitemOpen
  \bibfield  {author} {\bibinfo {author} {\bibfnamefont {C.~E.}\ \bibnamefont
  {Zachary}}\ and\ \bibinfo {author} {\bibfnamefont {S.}~\bibnamefont
  {Torquato}},\ }\bibfield  {title} {\enquote {\bibinfo {title}
  {Hyperuniformity in point patterns and two-phase heterogeneous media},}\
  }\href {https://doi.org/10.1088/1742-5468/2009/12/P12015} {\bibfield
  {journal} {\bibinfo  {journal} {J. Stat. Mech.: Theory \& Exp.}\ }\textbf
  {\bibinfo {volume} {2009}},\ \bibinfo {pages} {P12015} (\bibinfo {year}
  {2009})}\BibitemShut {NoStop}%
\bibitem [{\citenamefont {Torquato}(2021)}]{To21c}%
  \BibitemOpen
  \bibfield  {author} {\bibinfo {author} {\bibfnamefont {S.}~\bibnamefont
  {Torquato}},\ }\bibfield  {title} {\enquote {\bibinfo {title} {Structural
  characterization of many-particle systems on approach to hyperuniform
  states},}\ }\href {https://doi.org/10.1103/PhysRevE.103.052126} {\bibfield
  {journal} {\bibinfo  {journal} {Phys. Rev. E}\ }\textbf {\bibinfo {volume}
  {103}},\ \bibinfo {pages} {052126} (\bibinfo {year} {2021})}\BibitemShut
  {NoStop}%
\bibitem [{\citenamefont {Widom}(1965)}]{Wi65}%
  \BibitemOpen
  \bibfield  {author} {\bibinfo {author} {\bibfnamefont {B.}~\bibnamefont
  {Widom}},\ }\bibfield  {title} {\enquote {\bibinfo {title} {Equation of state
  in the neighborhood of the critical point},}\ }\href
  {https://doi.org/10.1063/1.1696618} {\bibfield  {journal} {\bibinfo
  {journal} {J. Chem. Phys.}\ }\textbf {\bibinfo {volume} {43}},\ \bibinfo
  {pages} {3898--3905} (\bibinfo {year} {1965})}\BibitemShut {NoStop}%
\bibitem [{\citenamefont {Kadanoff}(1966)}]{Ka66}%
  \BibitemOpen
  \bibfield  {author} {\bibinfo {author} {\bibfnamefont {L.~P.}\ \bibnamefont
  {Kadanoff}},\ }\bibfield  {title} {\enquote {\bibinfo {title} {Scaling laws
  for {I}sing models near {$T_c$}},}\ }\href
  {https://doi.org/10.1103/PhysicsPhysiqueFizika.2.263} {\bibfield  {journal}
  {\bibinfo  {journal} {Physics}\ }\textbf {\bibinfo {volume} {2}},\ \bibinfo
  {pages} {263--272} (\bibinfo {year} {1966})}\BibitemShut {NoStop}%
\bibitem [{\citenamefont {Fisher}(1967)}]{Fi67}%
  \BibitemOpen
  \bibfield  {author} {\bibinfo {author} {\bibfnamefont {M.~E.}\ \bibnamefont
  {Fisher}},\ }\bibfield  {title} {\enquote {\bibinfo {title} {The theory of
  equilibrium critical phenomena},}\ }\href
  {https://doi.org/10.1088/0034-4885/30/2/306} {\bibfield  {journal} {\bibinfo
  {journal} {Rep. Prog. Phys.}\ }\textbf {\bibinfo {volume} {30}},\ \bibinfo
  {pages} {615} (\bibinfo {year} {1967})}\BibitemShut {NoStop}%
\bibitem [{\citenamefont {Wilson}\ and\ \citenamefont {Kogut}(1974)}]{Wi74}%
  \BibitemOpen
  \bibfield  {author} {\bibinfo {author} {\bibfnamefont {K.~G.}\ \bibnamefont
  {Wilson}}\ and\ \bibinfo {author} {\bibfnamefont {J.}~\bibnamefont {Kogut}},\
  }\bibfield  {title} {\enquote {\bibinfo {title} {The renormalization group
  and the $\epsilon$ expansion},}\ }\href
  {https://doi.org/10.1016/0370-1573(74)90023-4} {\bibfield  {journal}
  {\bibinfo  {journal} {Phys. Rep.}\ }\textbf {\bibinfo {volume} {12}},\
  \bibinfo {pages} {75--199} (\bibinfo {year} {1974})}\BibitemShut {NoStop}%
\bibitem [{\citenamefont {Binney}\ \emph {et~al.}(1992)\citenamefont {Binney},
  \citenamefont {Dowrick}, \citenamefont {Fisher},\ and\ \citenamefont
  {Newman}}]{Bi92}%
  \BibitemOpen
  \bibfield  {author} {\bibinfo {author} {\bibfnamefont {J.~J.}\ \bibnamefont
  {Binney}}, \bibinfo {author} {\bibfnamefont {N.~J.}\ \bibnamefont {Dowrick}},
  \bibinfo {author} {\bibfnamefont {A.~J.}\ \bibnamefont {Fisher}},\ and\
  \bibinfo {author} {\bibfnamefont {M.~E.~J.}\ \bibnamefont {Newman}},\
  }\href@noop {} {\emph {\bibinfo {title} {The Theory of Critical Phenomena: An
  Introduction to the Renormalization Group}}}\ (\bibinfo  {publisher} {Oxford
  University Press},\ \bibinfo {address} {Oxford, England},\ \bibinfo {year}
  {1992})\BibitemShut {NoStop}%
\bibitem [{\citenamefont {O{\u g}uz}\ \emph {et~al.}(2019)\citenamefont {O{\u
  g}uz}, \citenamefont {Socolar}, \citenamefont {Steinhardt},\ and\
  \citenamefont {Torquato}}]{Og19}%
  \BibitemOpen
  \bibfield  {author} {\bibinfo {author} {\bibfnamefont {E.~C.}\ \bibnamefont
  {O{\u g}uz}}, \bibinfo {author} {\bibfnamefont {J.~E.~S.}\ \bibnamefont
  {Socolar}}, \bibinfo {author} {\bibfnamefont {P.~J.}\ \bibnamefont
  {Steinhardt}},\ and\ \bibinfo {author} {\bibfnamefont {S.}~\bibnamefont
  {Torquato}},\ }\bibfield  {title} {\enquote {\bibinfo {title}
  {Hyperuniformity and anti-hyperuniformity in one-dimensional substitution
  tilings},}\ }\href {https://doi.org/10.1107/S2053273318015528} {\bibfield
  {journal} {\bibinfo  {journal} {Acta Cryst. Section A: Foundations \&
  Advances}\ }\textbf {\bibinfo {volume} {A75}},\ \bibinfo {pages} {3--13}
  (\bibinfo {year} {2019})}\BibitemShut {NoStop}%
\bibitem [{\citenamefont {Torquato}(2018{\natexlab{b}})}]{To18b}%
  \BibitemOpen
  \bibfield  {author} {\bibinfo {author} {\bibfnamefont {S.}~\bibnamefont
  {Torquato}},\ }\bibfield  {title} {\enquote {\bibinfo {title} {Perspective:
  {B}asic understanding of condensed phases of matter via packing models},}\
  }\href@noop {} {\bibfield  {journal} {\bibinfo  {journal} {J. Chem. Phys.}\
  }\textbf {\bibinfo {volume} {149}},\ \bibinfo {pages} {020901} (\bibinfo
  {year} {2018}{\natexlab{b}})}\BibitemShut {NoStop}%
\bibitem [{\citenamefont {{Torquato}}, \citenamefont {{Zhang}},\ and\
  \citenamefont {{Stillinger}}(2015)}]{To15}%
  \BibitemOpen
  \bibfield  {author} {\bibinfo {author} {\bibfnamefont {S.}~\bibnamefont
  {{Torquato}}}, \bibinfo {author} {\bibfnamefont {G.}~\bibnamefont
  {{Zhang}}},\ and\ \bibinfo {author} {\bibfnamefont {F.~H.}\ \bibnamefont
  {{Stillinger}}},\ }\bibfield  {title} {\enquote {\bibinfo {title} {Ensemble
  theory for stealthy hyperuniform disordered ground states},}\ }\href@noop {}
  {\bibfield  {journal} {\bibinfo  {journal} {Phys. Rev. X}\ }\textbf {\bibinfo
  {volume} {5}},\ \bibinfo {pages} {021020} (\bibinfo {year}
  {2015})}\BibitemShut {NoStop}%
\bibitem [{\citenamefont {Koralov}(2005)}]{Ko05}%
  \BibitemOpen
  \bibfield  {author} {\bibinfo {author} {\bibfnamefont {L.}~\bibnamefont
  {Koralov}},\ }\bibfield  {title} {\enquote {\bibinfo {title} {The existence
  of pair potential corresponding to specified density and pair correlation},}\
  }\href {https://doi.org/10.1007/s11005-005-0343-9} {\bibfield  {journal}
  {\bibinfo  {journal} {Lett. Math. Phys.}\ }\textbf {\bibinfo {volume} {71}},\
  \bibinfo {pages} {135--148} (\bibinfo {year} {2005})}\BibitemShut {NoStop}%
\bibitem [{\citenamefont {Stell}(1977)}]{Stell77}%
  \BibitemOpen
  \bibfield  {author} {\bibinfo {author} {\bibfnamefont {G.}~\bibnamefont
  {Stell}},\ }\bibfield  {title} {\enquote {\bibinfo {title} {Fluids with
  long-range forces: {T}oward a simple analytic theory},}\ }in\ \href@noop {}
  {\emph {\bibinfo {booktitle} {Statistical Mechanics, Part A}}},\ \bibinfo
  {editor} {edited by\ \bibinfo {editor} {\bibfnamefont {B.~J.}\ \bibnamefont
  {Berne}}}\ (\bibinfo  {publisher} {Plenum Press},\ \bibinfo {address} {New
  York},\ \bibinfo {year} {1977})\ pp.\ \bibinfo {pages} {47--82}\BibitemShut
  {NoStop}%
\bibitem [{\citenamefont {Ornstein}\ and\ \citenamefont
  {Zernike}(1914)}]{Or14}%
  \BibitemOpen
  \bibfield  {author} {\bibinfo {author} {\bibfnamefont {L.~S.}\ \bibnamefont
  {Ornstein}}\ and\ \bibinfo {author} {\bibfnamefont {F.}~\bibnamefont
  {Zernike}},\ }\bibfield  {title} {\enquote {\bibinfo {title} {Accidental
  deviations of density and opalescence at the critical point of a single
  substance},}\ }\href@noop {} {\bibfield  {journal} {\bibinfo  {journal}
  {Proc. Akad. Sci. (Amsterdam)}\ }\textbf {\bibinfo {volume} {17}},\ \bibinfo
  {pages} {793--806} (\bibinfo {year} {1914})}\BibitemShut {NoStop}%
\bibitem [{\citenamefont {Yukawa}(1955)}]{Yu55}%
  \BibitemOpen
  \bibfield  {author} {\bibinfo {author} {\bibfnamefont {H.}~\bibnamefont
  {Yukawa}},\ }\bibfield  {title} {\enquote {\bibinfo {title} {{On the
  Interaction of Elementary Particles. I}},}\ }\href
  {https://doi.org/10.1143/PTPS.1.1} {\bibfield  {journal} {\bibinfo  {journal}
  {Prog. Theor. Phys. Suppl.}\ }\textbf {\bibinfo {volume} {1}},\ \bibinfo
  {pages} {1--10} (\bibinfo {year} {1955})}\BibitemShut {NoStop}%
\bibitem [{\citenamefont {Liu}\ and\ \citenamefont {Nocedal}(1989)}]{Liu89}%
  \BibitemOpen
  \bibfield  {author} {\bibinfo {author} {\bibfnamefont {D.~C.}\ \bibnamefont
  {Liu}}\ and\ \bibinfo {author} {\bibfnamefont {J.}~\bibnamefont {Nocedal}},\
  }\bibfield  {title} {\enquote {\bibinfo {title} {On the limited memory {BFGS}
  method for large scale optimization},}\ }\href
  {https://doi.org/https://doi.org/10.1007/BF01589116} {\bibfield  {journal}
  {\bibinfo  {journal} {Math. Programming}\ }\textbf {\bibinfo {volume} {45}},\
  \bibinfo {pages} {503--528} (\bibinfo {year} {1989})}\BibitemShut {NoStop}%
\bibitem [{\citenamefont {Hansen}(1973)}]{Ha73}%
  \BibitemOpen
  \bibfield  {author} {\bibinfo {author} {\bibfnamefont {J.~P.}\ \bibnamefont
  {Hansen}},\ }\bibfield  {title} {\enquote {\bibinfo {title} {Statistical
  mechanics of dense ionized matter. i. equilibrium properties of the classical
  one-component plasma},}\ }\href {https://doi.org/10.1103/PhysRevA.8.3096}
  {\bibfield  {journal} {\bibinfo  {journal} {Phys. Rev. A}\ }\textbf {\bibinfo
  {volume} {8}},\ \bibinfo {pages} {3096--3109} (\bibinfo {year}
  {1973})}\BibitemShut {NoStop}%
\bibitem [{\citenamefont {Gann}, \citenamefont {Chakravarty},\ and\
  \citenamefont {Chester}(1979)}]{Ga79}%
  \BibitemOpen
  \bibfield  {author} {\bibinfo {author} {\bibfnamefont {R.~C.}\ \bibnamefont
  {Gann}}, \bibinfo {author} {\bibfnamefont {S.}~\bibnamefont {Chakravarty}},\
  and\ \bibinfo {author} {\bibfnamefont {G.~V.}\ \bibnamefont {Chester}},\
  }\bibfield  {title} {\enquote {\bibinfo {title} {Monte {C}arlo simulation of
  the classical two-dimensional one-component plasma},}\ }\href
  {https://doi.org/10.1103/PhysRevB.20.326} {\bibfield  {journal} {\bibinfo
  {journal} {Phys. Rev. B}\ }\textbf {\bibinfo {volume} {20}},\ \bibinfo
  {pages} {326--344} (\bibinfo {year} {1979})}\BibitemShut {NoStop}%
\bibitem [{\citenamefont {Dyson}(1962)}]{Dy62a}%
  \BibitemOpen
  \bibfield  {author} {\bibinfo {author} {\bibfnamefont {F.~J.}\ \bibnamefont
  {Dyson}},\ }\bibfield  {title} {\enquote {\bibinfo {title} {Statistical
  theory of the energy levels of complex systems. {I}},}\ }\href
  {https://doi.org/10.1063/1.1703773} {\bibfield  {journal} {\bibinfo
  {journal} {J. Math. Phys.}\ }\textbf {\bibinfo {volume} {3}},\ \bibinfo
  {pages} {140--156} (\bibinfo {year} {1962})}\BibitemShut {NoStop}%
\bibitem [{\citenamefont {Ewald}(1921)}]{Ew21}%
  \BibitemOpen
  \bibfield  {author} {\bibinfo {author} {\bibfnamefont {P.~P.}\ \bibnamefont
  {Ewald}},\ }\bibfield  {title} {\enquote {\bibinfo {title} {Die berechnung
  optischer und elektrostatischer gitterpotentiale},}\ }\href
  {https://doi.org/https://doi.org/10.1002/andp.19213690304} {\bibfield
  {journal} {\bibinfo  {journal} {Annalen der Physik}\ }\textbf {\bibinfo
  {volume} {369}},\ \bibinfo {pages} {253--287} (\bibinfo {year}
  {1921})}\BibitemShut {NoStop}%
\bibitem [{Note5()}]{Note5}%
  \BibitemOpen
  \bibinfo {note} {Equations (\ref {ct_phic}), (\ref {cT_notPhic_alld}) and
  (\ref {kappa}) in the present work are Eqs. (118), (120) and (114) in Ref.
  \protect \citenum {To03a}, respectively, where $\xi =\kappa ^{-1}$. However,
  Eqs. (114) and (120) in Ref. \protect \citenum {To03a} contain trivial errors
  in constants due to a typo introduced in Eq. (111). The correct form of (111)
  in Ref. \protect \citenum {To03a} is \begin {equation} \protect \tilde
  {c}(k)=\protect \frac {-v_1(D)}{(1-\protect \frac {\phi }{\phi _c})+\protect
  \frac {1}{2(d+2)}(kD)^2+O((kD)^4)}. \end {equation} Equation (119) in Ref.
  \protect \citenum {To03a} is also incorrect due to this typo. The correct
  form of (119) is \begin {equation} c(r)= \begin {cases} 6\left (\protect
  \frac {\xi }{D}\right )\exp (-r/\xi ), \hskip 1em\relax d=1\\ 4 \ln \left
  (\protect \frac {r}{D}\right )\exp (-r/\xi ), \hskip 1em\relax d=2\\ -
  \protect \frac {2(d+2)}{d(d-2)} \left (\protect \frac {r}{D}\right
  )^{-(d-2)}\exp (-r/\xi ), \hskip 1em\relax d\geq 3. \end {cases} \label
  {ct_large_xi} \end {equation} We note that Eqs. (111), (114), (119), and
  (120) are the only equations affected by the typo. These errors do not affect
  the main conclusions of Ref. \protect \citenum {To03a}.}\BibitemShut {Stop}%
\bibitem [{\citenamefont {Wang}, \citenamefont {Stillinger},\ and\
  \citenamefont {Torquato}(2020)}]{Wa20}%
  \BibitemOpen
  \bibfield  {author} {\bibinfo {author} {\bibfnamefont {H.}~\bibnamefont
  {Wang}}, \bibinfo {author} {\bibfnamefont {F.~H.}\ \bibnamefont
  {Stillinger}},\ and\ \bibinfo {author} {\bibfnamefont {S.}~\bibnamefont
  {Torquato}},\ }\bibfield  {title} {\enquote {\bibinfo {title} {Sensitivity of
  pair statistics on pair potentials in many-body systems},}\ }\href
  {https://doi.org/10.1063/5.0021475} {\bibfield  {journal} {\bibinfo
  {journal} {J. Chem. Phys.}\ }\textbf {\bibinfo {volume} {150}},\ \bibinfo
  {pages} {124106} (\bibinfo {year} {2020})}\BibitemShut {NoStop}%
\bibitem [{\citenamefont {Torquato}, \citenamefont {Uche},\ and\ \citenamefont
  {Stillinger}(2006)}]{To06d}%
  \BibitemOpen
  \bibfield  {author} {\bibinfo {author} {\bibfnamefont {S.}~\bibnamefont
  {Torquato}}, \bibinfo {author} {\bibfnamefont {O.~U.}\ \bibnamefont {Uche}},\
  and\ \bibinfo {author} {\bibfnamefont {F.~H.}\ \bibnamefont {Stillinger}},\
  }\bibfield  {title} {\enquote {\bibinfo {title} {Random sequential addition
  of hard spheres in high {E}uclidean dimensions},}\ }\href
  {https://doi.org/10.1103/PhysRevE.74.061308} {\bibfield  {journal} {\bibinfo
  {journal} {Phys. Rev. E}\ }\textbf {\bibinfo {volume} {74}},\ \bibinfo
  {pages} {061308} (\bibinfo {year} {2006})}\BibitemShut {NoStop}%
\bibitem [{\citenamefont {Tomadakis}\ and\ \citenamefont
  {Robertson}(2003)}]{To03}%
  \BibitemOpen
  \bibfield  {author} {\bibinfo {author} {\bibfnamefont {M.~M.}\ \bibnamefont
  {Tomadakis}}\ and\ \bibinfo {author} {\bibfnamefont {T.~J.}\ \bibnamefont
  {Robertson}},\ }\bibfield  {title} {\enquote {\bibinfo {title} {Pore size
  distribution, survival probability, and relaxation time in random and ordered
  arrays of fibers},}\ }\href {https://doi.org/10.1063/1.1582431} {\bibfield
  {journal} {\bibinfo  {journal} {J. Chem. Phys.}\ }\textbf {\bibinfo {volume}
  {119}},\ \bibinfo {pages} {1741--1749} (\bibinfo {year} {2003})}\BibitemShut
  {NoStop}%
\bibitem [{\citenamefont {Stillinger}\ and\ \citenamefont
  {Weber}(1982)}]{St82a}%
  \BibitemOpen
  \bibfield  {author} {\bibinfo {author} {\bibfnamefont {F.~H.}\ \bibnamefont
  {Stillinger}}\ and\ \bibinfo {author} {\bibfnamefont {T.~A.}\ \bibnamefont
  {Weber}},\ }\bibfield  {title} {\enquote {\bibinfo {title} {Hidden structure
  in liquids},}\ }\href {https://doi.org/10.1103/PhysRevA.25.978} {\bibfield
  {journal} {\bibinfo  {journal} {Phys. Rev. A}\ }\textbf {\bibinfo {volume}
  {25}},\ \bibinfo {pages} {978--989} (\bibinfo {year} {1982})}\BibitemShut
  {NoStop}%
\bibitem [{\citenamefont {Maher}, \citenamefont {Stillinger},\ and\
  \citenamefont {Torquato}(2022)}]{Ma22a}%
  \BibitemOpen
  \bibfield  {author} {\bibinfo {author} {\bibfnamefont {C.~E.}\ \bibnamefont
  {Maher}}, \bibinfo {author} {\bibfnamefont {F.~H.}\ \bibnamefont
  {Stillinger}},\ and\ \bibinfo {author} {\bibfnamefont {S.}~\bibnamefont
  {Torquato}},\ }\bibfield  {title} {\enquote {\bibinfo {title}
  {Characterization of void space, large-scale structure, and transport
  properties of maximally random jammed packings of superballs},}\ }\href
  {https://doi.org/10.1103/PhysRevMaterials.6.025603} {\bibfield  {journal}
  {\bibinfo  {journal} {Phys. Rev. Mater.}\ }\textbf {\bibinfo {volume} {6}},\
  \bibinfo {pages} {025603} (\bibinfo {year} {2022})}\BibitemShut {NoStop}%
\bibitem [{\citenamefont {Kumar}\ \emph {et~al.}(2014)\citenamefont {Kumar},
  \citenamefont {Ray}, \citenamefont {Aswal},\ and\ \citenamefont
  {Kohlbrecher}}]{Ku14}%
  \BibitemOpen
  \bibfield  {author} {\bibinfo {author} {\bibfnamefont {S.}~\bibnamefont
  {Kumar}}, \bibinfo {author} {\bibfnamefont {D.}~\bibnamefont {Ray}}, \bibinfo
  {author} {\bibfnamefont {V.~K.}\ \bibnamefont {Aswal}},\ and\ \bibinfo
  {author} {\bibfnamefont {J.}~\bibnamefont {Kohlbrecher}},\ }\bibfield
  {title} {\enquote {\bibinfo {title} {Structure and interaction in the
  polymer-dependent reentrant phase behavior of a charged nanoparticle
  solution},}\ }\href {https://doi.org/10.1103/PhysRevE.90.042316} {\bibfield
  {journal} {\bibinfo  {journal} {Phys. Rev. E}\ }\textbf {\bibinfo {volume}
  {90}},\ \bibinfo {pages} {042316} (\bibinfo {year} {2014})}\BibitemShut
  {NoStop}%
\bibitem [{\citenamefont {Hansoge}\ \emph {et~al.}(2021)\citenamefont
  {Hansoge}, \citenamefont {Gupta}, \citenamefont {White}, \citenamefont
  {Giuntoli},\ and\ \citenamefont {Keten}}]{Ha21}%
  \BibitemOpen
  \bibfield  {author} {\bibinfo {author} {\bibfnamefont {N.~K.}\ \bibnamefont
  {Hansoge}}, \bibinfo {author} {\bibfnamefont {A.}~\bibnamefont {Gupta}},
  \bibinfo {author} {\bibfnamefont {H.}~\bibnamefont {White}}, \bibinfo
  {author} {\bibfnamefont {A.}~\bibnamefont {Giuntoli}},\ and\ \bibinfo
  {author} {\bibfnamefont {S.}~\bibnamefont {Keten}},\ }\bibfield  {title}
  {\enquote {\bibinfo {title} {Universal relation for effective interaction
  between polymer-grafted nanoparticles},}\ }\href
  {https://doi.org/10.1021/acs.macromol.0c02600} {\bibfield  {journal}
  {\bibinfo  {journal} {Macromolecules}\ }\textbf {\bibinfo {volume} {54}},\
  \bibinfo {pages} {3052--3064} (\bibinfo {year} {2021})}\BibitemShut {NoStop}%
\bibitem [{\citenamefont {Tucker}\ and\ \citenamefont {Maddox}(1998)}]{Tu98}%
  \BibitemOpen
  \bibfield  {author} {\bibinfo {author} {\bibfnamefont {S.~C.}\ \bibnamefont
  {Tucker}}\ and\ \bibinfo {author} {\bibfnamefont {M.~W.}\ \bibnamefont
  {Maddox}},\ }\bibfield  {title} {\enquote {\bibinfo {title} {The effect of
  solvent density inhomogeneities on solute dynamics in supercritical fluids: A
  theoretical perspective},}\ }\href {https://doi.org/10.1021/jp972382+}
  {\bibfield  {journal} {\bibinfo  {journal} {J. Phys. Chem. B}\ }\textbf
  {\bibinfo {volume} {102}},\ \bibinfo {pages} {2437--2453} (\bibinfo {year}
  {1998})}\BibitemShut {NoStop}%
\bibitem [{\citenamefont {Yu}\ \emph {et~al.}(2021)\citenamefont {Yu},
  \citenamefont {Qiu}, \citenamefont {Chong}, \citenamefont {Torquato},\ and\
  \citenamefont {Park}}]{Yu21}%
  \BibitemOpen
  \bibfield  {author} {\bibinfo {author} {\bibfnamefont {S.}~\bibnamefont
  {Yu}}, \bibinfo {author} {\bibfnamefont {C.-W.}\ \bibnamefont {Qiu}},
  \bibinfo {author} {\bibfnamefont {Y.}~\bibnamefont {Chong}}, \bibinfo
  {author} {\bibfnamefont {S.}~\bibnamefont {Torquato}},\ and\ \bibinfo
  {author} {\bibfnamefont {N.}~\bibnamefont {Park}},\ }\bibfield  {title}
  {\enquote {\bibinfo {title} {Engineered disorder in photonics},}\ }\href
  {https://doi.org/10.1038/s41578-020-00263-y} {\bibfield  {journal} {\bibinfo
  {journal} {Nature Rev. Mater.}\ }\textbf {\bibinfo {volume} {6}},\ \bibinfo
  {pages} {226--243} (\bibinfo {year} {2021})}\BibitemShut {NoStop}%
\bibitem [{\citenamefont {Wang}\ and\ \citenamefont
  {Torquato}(2022{\natexlab{b}})}]{Wa22a}%
  \BibitemOpen
  \bibfield  {author} {\bibinfo {author} {\bibfnamefont {H.}~\bibnamefont
  {Wang}}\ and\ \bibinfo {author} {\bibfnamefont {S.}~\bibnamefont
  {Torquato}},\ }\bibfield  {title} {\enquote {\bibinfo {title} {Dynamic
  measure of hyperuniformity and nonhyperuniformity in heterogeneous media via
  the diffusion spreadability},}\ }\href
  {https://doi.org/10.1103/PhysRevApplied.17.034022} {\bibfield  {journal}
  {\bibinfo  {journal} {Phys. Rev. Appl.}\ }\textbf {\bibinfo {volume} {17}},\
  \bibinfo {pages} {034022} (\bibinfo {year} {2022}{\natexlab{b}})}\BibitemShut
  {NoStop}%
\bibitem [{\citenamefont {Hejna}, \citenamefont {Steinhardt},\ and\
  \citenamefont {Torquato}(2013)}]{He13}%
  \BibitemOpen
  \bibfield  {author} {\bibinfo {author} {\bibfnamefont {M.}~\bibnamefont
  {Hejna}}, \bibinfo {author} {\bibfnamefont {P.~J.}\ \bibnamefont
  {Steinhardt}},\ and\ \bibinfo {author} {\bibfnamefont {S.}~\bibnamefont
  {Torquato}},\ }\bibfield  {title} {\enquote {\bibinfo {title} {Nearly
  hyperuniform network models of amorphous silicon},}\ }\href
  {https://doi.org/10.1103/PhysRevB.87.245204} {\bibfield  {journal} {\bibinfo
  {journal} {Phys. Rev. B}\ }\textbf {\bibinfo {volume} {87}},\ \bibinfo
  {pages} {245204} (\bibinfo {year} {2013})}\BibitemShut {NoStop}%
\bibitem [{\citenamefont {Atkinson}\ \emph {et~al.}(2016)\citenamefont
  {Atkinson}, \citenamefont {Zhang}, \citenamefont {Hopkins},\ and\
  \citenamefont {Torquato}}]{At16a}%
  \BibitemOpen
  \bibfield  {author} {\bibinfo {author} {\bibfnamefont {S.}~\bibnamefont
  {Atkinson}}, \bibinfo {author} {\bibfnamefont {G.}~\bibnamefont {Zhang}},
  \bibinfo {author} {\bibfnamefont {A.~B.}\ \bibnamefont {Hopkins}},\ and\
  \bibinfo {author} {\bibfnamefont {S.}~\bibnamefont {Torquato}},\ }\bibfield
  {title} {\enquote {\bibinfo {title} {Critical slowing down and
  hyperuniformity on approach to jamming},}\ }\href
  {https://doi.org/10.1103/PhysRevE.94.012902} {\bibfield  {journal} {\bibinfo
  {journal} {Phys. Rev. E}\ }\textbf {\bibinfo {volume} {94}},\ \bibinfo
  {pages} {012902} (\bibinfo {year} {2016})}\BibitemShut {NoStop}%
\bibitem [{\citenamefont {Xu}, \citenamefont {Douglas},\ and\ \citenamefont
  {Freed}(2016)}]{Xu16}%
  \BibitemOpen
  \bibfield  {author} {\bibinfo {author} {\bibfnamefont {W.-S.}\ \bibnamefont
  {Xu}}, \bibinfo {author} {\bibfnamefont {J.~F.}\ \bibnamefont {Douglas}},\
  and\ \bibinfo {author} {\bibfnamefont {K.~F.}\ \bibnamefont {Freed}},\
  }\bibfield  {title} {\enquote {\bibinfo {title} {Influence of cohesive energy
  on the thermodynamic properties of a model glass-forming polymer melt},}\
  }\href {https://doi.org/10.1021/acs.macromol.6b01503} {\bibfield  {journal}
  {\bibinfo  {journal} {Macromolecules}\ }\textbf {\bibinfo {volume} {49}},\
  \bibinfo {pages} {8341--8354} (\bibinfo {year} {2016})}\BibitemShut {NoStop}%
\bibitem [{\citenamefont {Chremos}\ and\ \citenamefont
  {Douglas}(2017)}]{Chr17}%
  \BibitemOpen
  \bibfield  {author} {\bibinfo {author} {\bibfnamefont {A.}~\bibnamefont
  {Chremos}}\ and\ \bibinfo {author} {\bibfnamefont {J.~F.}\ \bibnamefont
  {Douglas}},\ }\bibfield  {title} {\enquote {\bibinfo {title} {Particle
  localization and hyperuniformity of polymer-grafted nanoparticle
  materials},}\ }\href {https://doi.org/10.1002/andp.201600342} {\bibfield
  {journal} {\bibinfo  {journal} {Annalen der Physik}\ }\textbf {\bibinfo
  {volume} {529}} (\bibinfo {year} {2017}),\
  10.1002/andp.201600342}\BibitemShut {NoStop}%
\bibitem [{\citenamefont {{Martelli}}\ \emph {et~al.}(2017)\citenamefont
  {{Martelli}}, \citenamefont {{Torquato}}, \citenamefont {{Giovambattista}},\
  and\ \citenamefont {{Car}}}]{Mar17}%
  \BibitemOpen
  \bibfield  {author} {\bibinfo {author} {\bibfnamefont {F.}~\bibnamefont
  {{Martelli}}}, \bibinfo {author} {\bibfnamefont {S.}~\bibnamefont
  {{Torquato}}}, \bibinfo {author} {\bibfnamefont {N.}~\bibnamefont
  {{Giovambattista}}},\ and\ \bibinfo {author} {\bibfnamefont {R.}~\bibnamefont
  {{Car}}},\ }\bibfield  {title} {\enquote {\bibinfo {title} {Large-scale
  structure and hyperuniformity of amorphous ices},}\ }\href
  {https://doi.org/10.1103/PhysRevLett.119.136002} {\bibfield  {journal}
  {\bibinfo  {journal} {Phys. Rev. Lett.}\ }\textbf {\bibinfo {volume} {119}},\
  \bibinfo {pages} {136002} (\bibinfo {year} {2017})}\BibitemShut {NoStop}%
\end{thebibliography}
%

\end{document}